\documentclass[12pt]{iopart}
\usepackage{color}

\usepackage{enumitem}
\usepackage{graphicx,amssymb,bm}
\usepackage[caption=false]{subfig}
\usepackage{float}
\usepackage{mathscinet}
\usepackage{cite}

\begin{document}

\title[]{Opinion formation on social networks with algorithmic bias: Dynamics and bias imbalance}
\author{Antonio F. Peralta$^1$, J\'anos Kert\'esz$^1$$^{,}$$^2$, and Gerardo I\~{n}iguez$^1$$^{,}$$^3$$^{,}$$^4$}
\address{$^1$ Department of Network and Data Science, Central European University, A-1100 Vienna, Austria}
\address{$^2$ Complexity Science Hub, A-1080 Vienna, Austria}
\address{$^3$ Department of Computer Science, Aalto University School of Science, FI-00076 Aalto, Finland}
\address{$^4$ Centro de Ciencias de la Complejidad, Universidad Nacional Auton\'{o}ma de M\'{e}xico, 04510 Ciudad de M\'{e}xico, Mexico}

\vspace{10pt}
\begin{indented}
\item[]\today
\end{indented}

\begin{abstract}
We investigate opinion dynamics and information spreading on networks under the influence of content filtering technologies. The filtering mechanism, present in many online social platforms, reduces individuals' exposure to disagreeing opinions, producing algorithmic bias. We derive evolution equations for global opinion variables in the presence of algorithmic bias, network community structure, noise (independent behavior of individuals), and pairwise or group interactions. We consider the case where the social platform shows a predilection for one opinion over its opposite, unbalancing the dynamics in favor of that opinion. We show that if the imbalance is strong enough, it may determine the final global opinion and the dynamical behavior of the population. We find a complex phase diagram including phases of coexistence, consensus, and polarization of opinions as possible final states of the model, with phase transitions of different order between them.
The fixed point structure of the equations determines the dynamics to a large extent. We focus on the time needed for convergence and conclude that this quantity varies within a wide range, showing occasionally signatures of critical slowing down and meta-stability.
\end{abstract}

\maketitle

\section{Introduction}

The collective behavior of a system made of interacting individuals can be successfully analyzed using agent-based models, as shown in many examples across various disciplines \cite{Castellano:2009a,Satorras:2015,Porter:2016}. In these models, individuals (or agents) are often pictured as nodes in a network \cite{Newman:2003b,Lambiotte:2021}, where the links represent the possible interactions between them. Each node holds a dynamical state variable whose interpretation depends on the context of the model. In opinion dynamics this opinion  variable \cite{Castellano:2009a} can be considered as the political party preference of an individual (e.g., liberal or conservative), her inclination towards or against some regulation or initiative, etc. The usefulness of the approach lies in the simplicity of the setting, leading, nevertheless, to complex phenomena due to collective effects. The possibility of considering various elements that are hypothesized to be relevant for opinion formation (such as structural and dynamical heterogeneities) deeply improves our understanding of the underlying social phenomena.

In recent years, human communication has changed dramatically, and moved from traditional media (face-to-face, phone, or mass media like television and the press) to online social media platforms (Google, Twitter, Facebook, etc.) \cite{Lazer:2009,Conte:2012}. In contrast to earlier media channels, online social networks control the information that users see and send to each other by means of personalized filtering algorithms \cite{Nikolov:2018}. These algorithms record individual information about users' preferences and then filter incoming data accordingly \cite{Bozdag:2013, Moller:2018}. Hence, people tend to be exposed to opinions they already agree with, producing so-called \emph{algorithmic bias}. This reinforcement feature changes, ultimately, the global behavior of the population \cite{DelVicario:2016,Bail:2018,Ciampaglia:2018,Blex:2020}, promoting phenomena like ``filter bubbles" or ``echo chambers", where people divide in groups with opposing views that barely interact with each other. Explaining how and when the polarization of opinion groups emerges within a population is of crucial importance \cite{Baumann:2020,Cinelli:2021}. It is of particular interest to understand how copying or herding processes (typical signatures of human social behavior), when coupled to algorithmic bias, may enhance or decrease polarization. These two ingredients can be implemented in the formalism of agent-based models, leading to a flexible mathematical framework open to analytical and numerical treatments. Ultimately, the results of the models can be interpreted to help devise potential strategies to mitigate the negative effects of algorithmic bias.

Previous modeling efforts \cite{Sirbu:2019} have been made to consider algorithmic bias in bounded confidence models \cite{Deffuant:2000}, where the opinion variable of individuals is a real continuous variable on an interval. The filtering algorithm requires the opinions of two individuals to be similar enough to be able to interact, and bias means that similar people with similar opinions have a greater chance to meet, leading to enhanced polarization and fragmentation in opinion space. Another class of models considers opinions to be discrete (a binary variable in the simplest case) \cite{Gleeson:2011,Gleeson:2013}. Perra and Rocha \cite{Perra:2019} have studied algorithmic bias in such a model by considering that the opinion of an individual is influenced by its neighbors in the network, and filtered in various ways. An alternative implementation of algorithmic bias in binary-state models has been proposed in \cite{DeMarzo:2020}. In this case, the social platform records information about all the previous opinions of individuals, and then influences them to keep the opinion that has been held for the longest time, similarly to a memory or `inertia' effect \cite{Stark:2008,Peralta:2019,Peralta:2020_mem}. All these implementations of algorithmic bias in opinion dynamics modeling suggest that polarization is a consequence of both the social behavior of individuals and the content filtering algorithms constraining their actions.

In the present work, similarly to the approach of \cite{Perra:2019}, we consider that a fraction of the neighbors of an individual holding disagreeing opinions are filtered, and thus interactions with those neighbors are not possible. Recently we have proposed a general formalism within the binary-state approach that includes this implementation of algorithmic bias  \cite{Peralta:2021}. We have extended previous theoretical tools to describe the macroscopic dynamics on networks, including mean-field and pair approximations. We have also explored modular community structures, a crucial ingredient to characterize opinion polarization, i.e. the division of the population into opinion groups. We have studied the static, asymptotic behavior of archetypal models of opinion formation in the presence of algorithmic bias and concluded that, systematically, pairwise interactions lead to polarization, while group interactions promote coexistence of opinions. Here we use the same formalism of \cite{Peralta:2021}, but focus on dynamical aspects of the opinion formation process. We also extend the algorithmic bias mechanism to consider situations in which the online social platform promotes one opinion over the other, thus unbalancing the dynamics in favor of the preferred opinion by the platform, in what we call \emph{bias asymmetry}.

We use a prototypical model of social behavior, the language model \cite{Abrams:2003, Vazquez:2010}, which takes into account both pairwise and group-based (copying) interactions depending on the value of a tunable parameter $\alpha$. We also include the possibility that individuals act independently of their neighbors \cite{Kirman:1993, Granovsky:1995, Peralta:2018}, which we denote as noise with intensity $Q$. This general framework enables us to consider several interaction mechanisms and leads to various opinion formation scenarios. In \cite{Peralta:2021}, we have considered other archetypal models of opinion formation (including voter-like and majority-vote models), and realized that the language model essentially interpolates between dynamics with either pairwise or group interactions depending on the value of $\alpha$, and is thus a good candidate to explore the effects of bias asymmetry within a single model. 

The paper is organized as follows. In Sec. \ref{sec_model_def} we define the opinion formation model, algorithmic bias, and social network with community structure we use. In Sec. \ref{sec_mean_field} we derive a set of mean-field equations that describe the global dynamics of the model, and derive its associated fixed points (stationary states of the dynamics). In Sec. \ref{sec_local} we explore the local dynamics and stability of the fixed points and build the phase diagram of the model. In Sec. \ref{sec_metastable} we present a detailed study of temporal behavior of the considered opinion dynamics model using numerical simulations and some theoretical tools, with particular emphasis on the role of initial conditions, and the behavior of the time to reach the final state. Throughout this paper we will pay special attention to the effect of algorithmic bias and its asymmetry.

\section{Model and definitions}\label{sec_model_def}

We consider the formalism of binary-state dynamics \cite{Gleeson:2011, Gleeson:2013} as basic ground for modeling opinion formation. The model system is composed of a set of $i=1, \dots, N$ individuals, each one holding a binary-state (opinion) variable $s_{i}(t) = 0, 1$ at time $t$ (e.g., liberal or conservative in a political setting). We define the macroscopic state (global opinion) of the system as $\rho=N^{-1} \sum_{i=1}^{N} s_{i} \in [0, 1]$, i.e. the density of individuals in state $1$. Individuals are represented by nodes of an (undirected) network, and links in the network correspond to some social relationship between them, such that the opinion of an individual can be influenced by its neighbors in the network. The state of node $i$ changes according to rates depending on the specific dynamics and the network structure: with {\it ``infection"} rate $F_{k_{i}, m_{i}}$ from $s_{i}=0 \rightarrow 1$, and with {\it ``recovery"} rate $R_{k_{i}, m_{i}}$ from $s_{i}=1 \rightarrow 0$, where $k_{i}$ is the degree of the node in the network and $m_{i} \in [0, k_{i}]$ is the number of (nearest) neighbors of $i$ in state $s_{j}=1$. (Here the names of the rates refer to the analogy with epidemic spreading.)

In the following we specify the spreading dynamics, i.e. the functional form of the rates $F_{k,m}$ and $R_{k,m}$, and the network structure. As for the spreading dynamics, we incorporate algorithmic bias (representing content filtering as implemented in many online social platforms) that influences and controls the way people interact. For the network structure, we include modules or communities by dividing the population into groups with tunable connectivity.

\subsection{Transition rates in the language model}

As mentioned above, we focus on the \emph{language model} \cite{Abrams:2003,Vazquez:2010}, which is able to describe both pairwise and group interactions. The transition rates are as follows:

\begin{eqnarray}
\label{infection_language}
F_{k,m} &=& Q + (1-2 Q) \left(\frac{m}{k} \right)^{\alpha},\\
\label{recovery_language}
R_{k,m} &=& Q + (1-2 Q) \left(\frac{k-m}{k} \right)^{\alpha},
\end{eqnarray} 
with $Q \in [0,1/2]$ and $\alpha \in (0, \infty)$ as tuning parameters. The model takes into account two mechanisms driving the dynamics: (i) noisy or idiosyncratic changes of state, with intensity $Q$; and (ii) herding or copying the states of neighbors with probability proportional to the fraction of neighbors in the opposite state to a power $\alpha$. The rates in Eqs. (\ref{infection_language}, \ref{recovery_language}) were first studied in the case without noise ($Q=0$) to model the dynamics of language death \cite{Abrams:2003}, but the same model has been applied to other types of social human behavior, for example in the context of opinion formation in social media \cite{Xiong:2014}. The role of noise ($Q>0$) has been extensively studied of late, as for the \emph{non-linear noisy voter model} in \cite{Peralta:2018} with $\alpha$ a real (continuous) number, and for the \emph{q-voter model} \cite{Castellano:2009b,Nyczka:2012,Jedrzejewski:2017} with $\alpha=q$ a positive integer. The language model encapsulates a wide variety of phenomena depending on the value of $\alpha$. E.g., for the mean field (complete network) the following regions can be distinguished (i) low $0 < \alpha < 2$ (\emph{pairwise} interactions); (ii) high $2 < \alpha < 5$, (\emph{group} interactions); and (iii) very high $\alpha > 5$. Each of these cases displays a distinct phenomenology \cite{Peralta:2021,Peralta:2018} and represents a different archetypal way for humans to influence each other (either in pairs or in groups).

Note that the original rates in Eqs. (\ref{infection_language}, \ref{recovery_language}) fulfill the ``up-down symmetry" condition $R_{k,m} = F_{k,k-m}$ but, as we will show next, Eqs. (\ref{efective_rate_def_infection}, \ref{efective_rate_def_recovery}) are only symmetric for $b_{0} = b_{1}$. In other words, unbalanced algorithmic bias breaks the symmetry of the system and favors one opinion over the other.

\subsection{Algorithmic bias}

A simple implementation of algorithmic bias has been proposed by us in a previous study \cite{Peralta:2021}. Here we generalize the definition of that paper by introducing two parameters characterizing the bias, instead of one. The \emph{bias intensities} $b_{0}$ and $b_{1}$ (where the subscripts refer to state $0$ or $1$) take values in the interval $[0,1]$. These parameters measure the probabilities that the online platform filters out a neighbor in the opposite state, $0$ or $1$, (disagreeing opinion) of an individual with a given opinion, so that further interactions with that neighbor cannot take place. This mechanism of content filtering can be implemented in the formalism of any rate governed binary-state model by considering the following effective transition rates $F_{k,m}^{*}$, $R_{k,m}^{*}$:

\begin{eqnarray}
\label{efective_rate_def_infection}
F_{k,m}^{*}(b_{1}) &=& \sum_{i=0}^{m} B_{m,i}(1-b_{1}) F_{k-m+i,i},\\
\label{efective_rate_def_recovery}
R_{k,m}^{*}(b_{0}) &=& \sum_{s=0}^{k-m} B_{k-m,s}(1-b_{0}) R_{m+s,m},
\end{eqnarray}
with the binomial $B_{k,m}(1-b)= {{k}\choose{m}} (1-b)^{m} b^{k-m}$. Eqs. (\ref{efective_rate_def_infection}, \ref{efective_rate_def_recovery}) express the average rates of changing state, after removing with probability $b_{0}$ or $b_{1}$ a subset of neighbors in the opposite state ($b_{0}$ if neighbors are in state $s=0$ and $b_{1}$ for $s=1$). We define the \emph{total bias intensity} $b = (b_{0}+b_{1})/2$, and the \emph{bias asymmetry} $\Delta b = b_{1} - b_{0}$, with $b_{0} = b - \Delta b/2$ and $b_{1} = b + \Delta b/2$. For $\Delta b > 0$ the social platform favors $s=0$, while for $\Delta b < 0$ it favors $s=1$. In \cite{Peralta:2021} we have implemented bias in a similar way to Eqs. (\ref{efective_rate_def_infection}, \ref{efective_rate_def_recovery}), but with $b_{0} = b_{1} = b$ (i.e. $\Delta b=0$), such that the role of algorithmic bias is symmetric, or balanced across opinions.

\subsection{Modular structure}

The social network of interactions between individuals is fully specified by the adjacency matrix $A_{ij}$, with elements equal to $1$ if $i$ and $j$ are connected and $0$ otherwise. For simplicity we consider the degree distribution $P_{k}$ as the only relevant structural feature of the network, with average degree $z = \sum_{k} P_{k} k$. We use the standard configuration model \cite{Catanzaro:2005} to produce synthetic networks in the corresponding numerical simulations. 

When the network displays modular (community) structure, we can divide the population into  groups with higher connectivity to nodes inside the group than to those outside. Assuming for simplicity two modules, nodes $i=1, \dots, N_{1}$ are in group 1 of size $N_{1}$, and nodes $i=N_{1}+1, \dots, N_{1}+N_{2}=N$ in group 2 of size $N_{2}$. The two groups have different connectivity depending on whether links join nodes of the same group or between groups. In this way, two nodes in the same group are more likely to be connected than if they belong to different groups. This group {\it homophily} \cite{McPherson:2001,Asikainen:2020} is a property that many real-world social networks show and is strongly correlated with the opinion of individuals. In other words, the opinion of an individual is ultimately related to the group it belongs to, in what we call a polarization state, or homophily for those who share the same ideas as one. In order to characterize the macroscopic state accordingly, we need at least two variables: $\rho_{1} =N_{1}^{-1} \sum_{i=1}^{N_{1}} s_{i} \in [0,1]$ and $\rho_{2} =N_{2}^{-1} \sum_{i=N_{1}+1}^{N_{2}} s_{i} \in [0,1]$, the density of nodes in state $1$ in groups $1$ and $2$, respectively, with total $\rho=\frac{N_{1}}{N} \rho_{1} + \frac{N_{2}}{N} \rho_{2} \in [0,1]$. Additionally, we define \emph{polarization} as $P=\vert \rho_{1} - \rho_{2}\vert \in [0,1]$, which measures the degree of opinion dissimilarity between groups.

We consider four average degrees, $z_{1}$, $z_{12}$, $z_{21}$, and $z_{2}$, defining the connectivity inside and between groups. Parameter $z_{1}$ ($z_{2}$) is the average degree only considering links that join nodes within group $1$ ($2$), while $z_{12}$ ($z_{21}$) is the average degree only considering links that depart from group $1$ and end up in group $2$ (from $2$ to $1$). The total number of links that go from group $1$ to $2$ is the same as those that go from $2$ to $1$, so we have the constraint $N_{1} z_{12} = N_{2} z_{21}$.


\section{Mean-field description}\label{sec_mean_field}

We derive a set of mean-field evolution equations for the two macroscopic variables $\rho_{1}(t)$ and $\rho_{2}(t)$, of general validity in the thermodynamic ($N \rightarrow \infty$) and highly connected ($z_{1}, z_{12}, z_{21}, z_{2} \rightarrow \infty$) limit, with constant ratios of the average degrees. Even in a finite system with high connectivity, the mean field description is a good approximation of the dynamics, and it captures the phenomenology of the model well. For large values of $N$, the opinion variables fluctuate slightly around their average values, i.e., $\rho_{1}(t) \approx \langle \rho_{1}(t) \rangle$, $\rho_{2}(t) \approx \langle \rho_{2}(t) \rangle$. Thus, throughout the following mean-field description, $\rho_{1}(t)$ and $\rho_{2}(t)$ refer to average values over realizations.

In order to derive the mean-field equations \cite{Peralta:2021} we first define the average rate of changing state \cite{Gleeson:2013} as

\begin{equation}
\label{mean_field_rate}
f[x] \equiv \sum_{k} \frac{P_{k} k}{z} \sum_{m=0}^{k} F_{k,m} B_{k,m}(x),
\end{equation}
where $x$ is the probability of finding a neighbor in state $1$, and $P_{k} k / z$ is the probability that a link connects to a node with degree $k$. In order to obtain a closed description of the dynamics, we must relate the probability $x$ to the description variables $\rho_{1}$ and $\rho_{2}$. The probability $x$ depends on the group to which the node belongs. In the case of group $1$ we have

\begin{equation}
\label{prob_community}  
x_{1}=\frac{N_{1} z_{1} \rho_{1} + N_{2} z_{21} \rho_{2}}{N_{1} z_{1} + N_{2} z_{21}} = \frac{ \rho_{1} + p_{1} \rho_{2}}{1 + p_{1}},
\end{equation}
with $p_{1}=N_{2} z_{21}/N_{1} z_{1}=z_{12}/z_{1}$, and, similarly, for group $2$ exchanging the labels $1 \leftrightarrow 2$, with $p_{2}=N_{1} z_{12}/N_{2} z_{2} = z_{21}/z_{2}$. Eq. (\ref{prob_community}) is the ratio of the number of links coming out of nodes in state $1$ that connect to nodes in group $1$, and the number of links coming out of nodes in group $1$.

If we consider algorithmic bias ($b_{1}>0$), we must use the effective $F^{*}_{k,m}$ of Eq. (\ref{efective_rate_def_infection}) instead of $F_{k,m}$ to calculate the average rate $f^{*}[x]$ in the network using Eq. (\ref{mean_field_rate}). Applying an argument based on the highly connected limit ($z \rightarrow \infty$) \cite{Peralta:2021}, the effective average rate with bias reduces to

\begin{equation}
\label{efective_mean_field_rate_inf}
f^{*}[x] \approx f \left[\frac{(1-b_{1})x}{1-b_{1} x} \right].
\end{equation}
When $b_{1}=0$ we recover $f^{*}[x]=f[x]$. An analogous procedure can be applied to the effective recovery rate $R^{*}_{k,m}$ of Eq. (\ref{efective_rate_def_recovery}), leading to $r^{*}[x]$, which in the presence of up-down symmetry ($R_{k,m}=F_{k,k-m}$) is given by

\begin{equation}
\label{efective_mean_field_rate_rec}
r^{*}[x] \approx f \left[\frac{(1-b_{0})(1-x)}{1-b_{0} (1-x)} \right].
\end{equation}
After defining the effective average rates $f^{*}[x]$, $r^{*}[x]$ and the probabilities $x_{1,2}$, we obtain a system of two differential (mean-field) equations for the dynamics of the state variables $\vec{\rho}(t)$ with components $\rho_{1}(t)$ and $\rho_{2}(t)$:
\begin{eqnarray}
\label{dyn_rho1}
\frac{d\rho_{1}}{dt} &=& (1-\rho_{1}) f \left[ \frac{(1-b_{1})(\rho_1 + p_{1} \rho_2)}{1+p_{1}- b_{1} (\rho_1 + p_{1} \rho_2)} \right] \nonumber\\
&-& \rho_{1} f \left[ \frac{(1-b_{0})(1-\rho_1 + p_{1} (1-\rho_2))}{1+p_{1}-b_{0} (1-\rho_1 + p_{1} (1-\rho_2))} \right] \equiv  \mu_{1}[\rho_{1},\rho_{2}],\\
\label{dyn_rho2}
\frac{d\rho_{2}}{dt} &=& (1-\rho_{2}) f \left[ \frac{(1-b_{1})(\rho_2 + p_{2} \rho_1)}{1+p_{2}-b_{1}(\rho_2 + p_{2} \rho_1)} \right] \nonumber\\
&-& \rho_{2} f \left[ \frac{(1-b_{0})(1-\rho_2 + p_{2} (1-\rho_1))}{1+p_{2}-b_{0}(1-\rho_2 + p_{2} (1-\rho_1))} \right] \equiv \mu_{2}[\rho_{1},\rho_{2}].
\end{eqnarray}
This mean-field description of the opinion formation model has the social behavioral parameters $(Q, \alpha)$, the algorithmic bias parameters $(b, \Delta b)$, and the group connectivity parameters $(p_{1},p_{2})$, together with the initial condition $\rho_{1}(0)$, $\rho_{2}(0)$.

\subsection{Fixed point structure and stationary solutions }

\begin{figure}[ht]
\centering
\includegraphics[width=\textwidth]{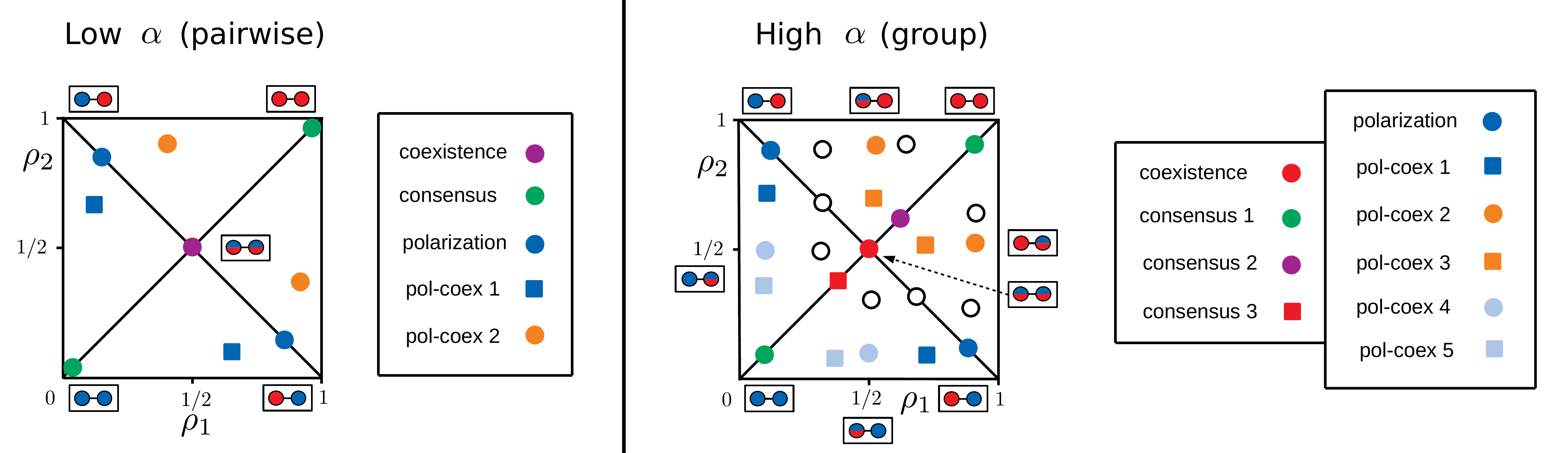}
\caption{Schematic representation of fixed points of opinion dynamics in ($\rho_{1}$, $\rho_{2}$)-phase space. The left (right) side corresponds to the low (high) $\alpha$ regime of the model. Joined colored circles inside a square represent the opinion states $s=0$ (blue) or $s=1$ (red) of each group (e.g., two circles of the same color represent consensus, while circles of different colors represent polarization). The legend displays the color coding and names of the most relevant fixed points. Pairs of fixed points [circle (stable) and square (unstable)] have the same color if they disappear/appear together for specific values of the parameters.}
\label{fig:color_scheme}
\end{figure}

The stationary results of Eqs. (\ref{dyn_rho1}, \ref{dyn_rho2}), i.e. the infinite time limit $\rho_{1}(\infty)$ and $\rho_{2}(\infty)$, correspond to the stable fixed points. The fixed points $\rho^{{\rm st}}_{1}$, $\rho^{{\rm st}}_{2}$ are obtained from the condition
\begin{eqnarray}
\label{fixed_points_pol1}
\mu_{1}[\rho^{{\rm st}}_{1},\rho^{{\rm st}}_{2}] &=& 0,\\ \label{fixed_points_pol2}
\mu_{2}[\rho^{{\rm st}}_{1},\rho^{{\rm st}}_{2}] &=& 0.
\end{eqnarray}
The study of the solutions of Eqs. (\ref{fixed_points_pol1}, \ref{fixed_points_pol2}) as a function of the parameters is a first step in understanding the dynamics and general behavior of the model. In Fig. \ref{fig:color_scheme} we show the positions in ($\rho_{1}$, $\rho_{2}$)-phase space of all possible fixed points in the model, together with color and name coding to identify and refer to them easily in the following sections and figures. Note the presence of the collective opinion states most relevant to our discussion: \emph{consensus} ($\rho^{{\rm st}}_{1} = \rho^{{\rm st}}_{2} \approx 0$ or $\rho^{{\rm st}}_{1} = \rho^{{\rm st}}_{2} \approx 1$), \emph{coexistence} ($\rho^{{\rm st}}_{1} = \rho^{{\rm st}}_{2} \approx 1/2$),  and \emph{polarization} ($\rho^{{\rm st}}_{1} \approx 0$, $\rho^{{\rm st}}_{2} \approx 1$ or $\rho^{{\rm st}}_{1} \approx 1$, $\rho^{{\rm st}}_{2} \approx 0$). All these states are possible in the low (pair) and high (group) $\alpha$ regimes for certain values of the model parameters $(Q, \alpha, b, \Delta b, p_{1}, p_{2})$.

\section{Local dynamics and stability}\label{sec_local}

A basic step in understanding the dynamics of the model is to explore the vector fields of Eqs. (\ref{dyn_rho1}, \ref{dyn_rho2}) and the associated trajectories close to the fixed points $\rho_{1}(t) \approx \rho^{{\rm st}}_{1}$, $\rho_{2}(t) \approx \rho^{{\rm st}}_{2}$. This can be done by means of a linearization of the dynamical equations. The linearization process leads to an exponential solution,
\begin{eqnarray}
\label{rho1_lin}
\rho_{1}(t) &\approx& \rho^{{\rm st}}_{1}+ C_{1} v_{11} e^{-\lambda_{1} t} + C_{2} v_{21} e^{-\lambda_{2} t},\\
\label{rho2_lin}
\rho_{2}(t) &\approx& \rho^{{\rm st}}_{2}+C_{1} v_{12} e^{-\lambda_{1} t} + C_{2} v_{22} e^{-\lambda_{2} t},
\end{eqnarray}
where $\vec{v}_{1}=\left( v_{11} , v_{12} \right) $ and $\vec{v}_{2}= \left( v_{21} , v_{22} \right) $ are the eigenvectors with associated eigenvalues $\lambda_{1}$ and $\lambda_{2}$ of the Jacobian matrix $J_{ij}=\frac{\partial \mu_{i}}{\partial \rho_{j}}$, evaluated at the fixed point $\rho^{{\rm st}}_{1}$, $\rho^{{\rm st}}_{2}$. $C_{1}$ and $C_{2}$ can be calculated from the initial condition as
\begin{eqnarray}
C_{1} &=& \frac{v_{21} \rho_{2}^{*}-v_{22} \rho_{1}^{*}}{v_{12}v_{21}-v_{11}v_{22}},\\
C_{2} &=& \frac{v_{12} \rho_{1}^{*}-v_{11} \rho_{2}^{*}}{v_{12}v_{21}-v_{11}v_{22}},
\end{eqnarray}
with $\rho_{1}^{*}=\rho_{1}(0)-\rho^{{\rm st}}_{1}$ and $\rho_{2}^{*}=\rho_{2}(0)-\rho^{{\rm st}}_{2}$. 

The stability of a fixed point is determined by the sign of (the real part of) the associated eigenvalues: for $\lambda_{1,2}<0$ it is stable, for $\lambda_{1,2}>0$ it is unstable, and for $\lambda_{1}<0$, $\lambda_{2}>0$ or $\lambda_{1}>0$, $\lambda_{2}<0$ it is a saddle point. Only when the fixed point is stable, we expect trajectories to converge to the fixed point in the long time limit [$\rho_{1}(t) \rightarrow \rho^{{\rm st}}_{1}$, $\rho_{2}(t) \rightarrow \rho^{{\rm st}}_{2}$].

The eigenvalues can be calculated as
\begin{equation}
\lambda_{1,2}=\frac{1}{2} \left( \tau \pm \sqrt{\tau^2-4 \Delta} \right), \hspace{0.4cm} \Delta=\lambda_{1} \lambda_{2}, \hspace{0.4cm} \tau=\lambda_{1} + \lambda_{2},
\end{equation}
where $\tau = J_{11} + J_{22}$ is the trace and $\Delta = J_{11} J_{22} - J_{12} J_{21}$ the determinant of the Jacobian matrix evaluated at the fixed point $\rho^{{\rm st}}_{1}$, $\rho^{{\rm st}}_{2}$. If $\tau^2 > 4 \Delta$ the eigenvalues only have a real part (the case for all fixed points in Fig. \ref{fig:color_scheme}). The condition for a transition (bifurcation) is that one of the eigenvalues becomes zero (a so-called marginal stability), or equivalently $\Delta[\rho^{{\rm st}}_{1},\rho^{{\rm st}}_{2}]=0$. This condition together with Eqs. (\ref{fixed_points_pol1}, \ref{fixed_points_pol2}) determines the transition lines and the phase diagrams. At a transition we expect some fixed points to appear or disappear (usually in couples).

\begin{figure}[ht]
\centering
\subfloat[]{\label{fig:phases_pair:a}\includegraphics[width=0.4\textwidth]{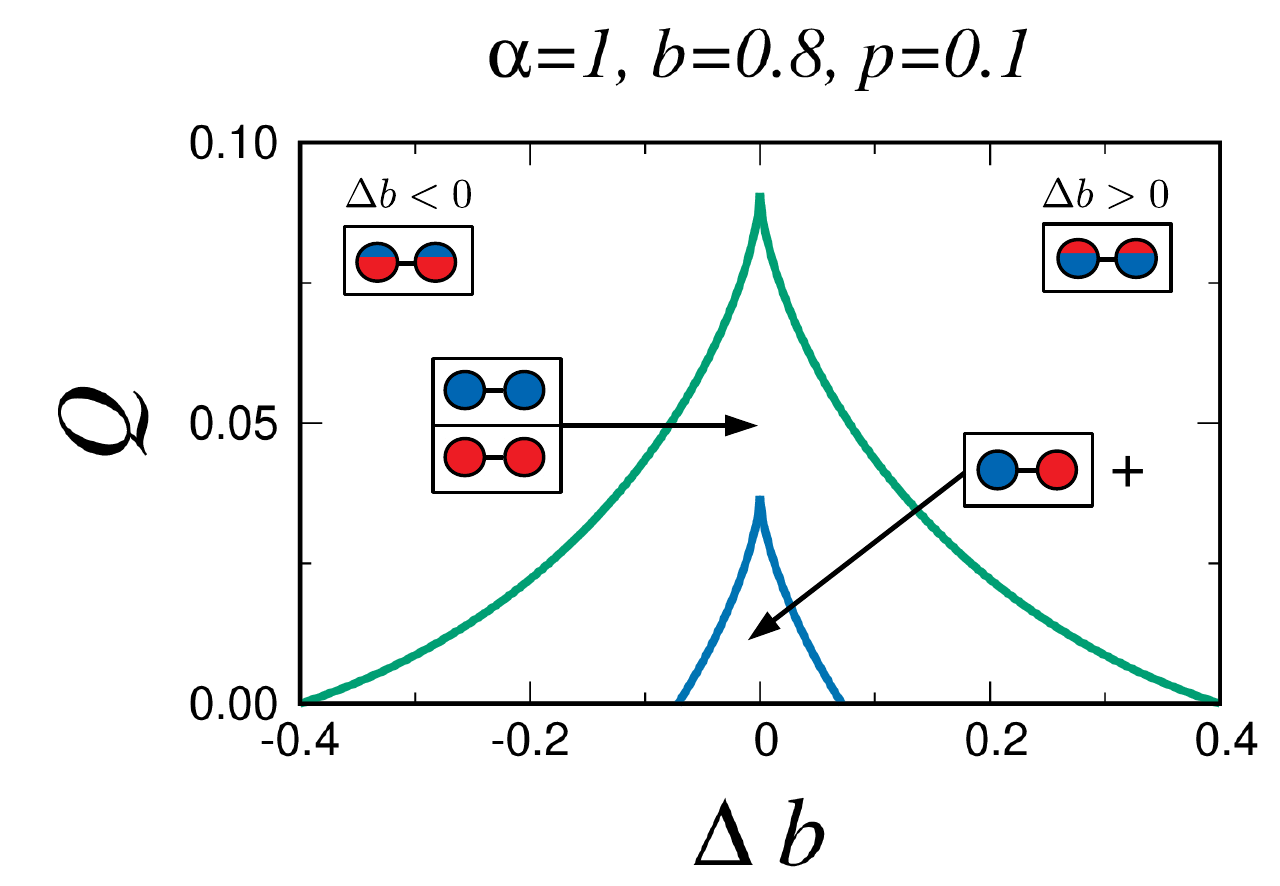}}
\subfloat[]{\label{fig:phases_pair:b}\includegraphics[width=0.4\textwidth]{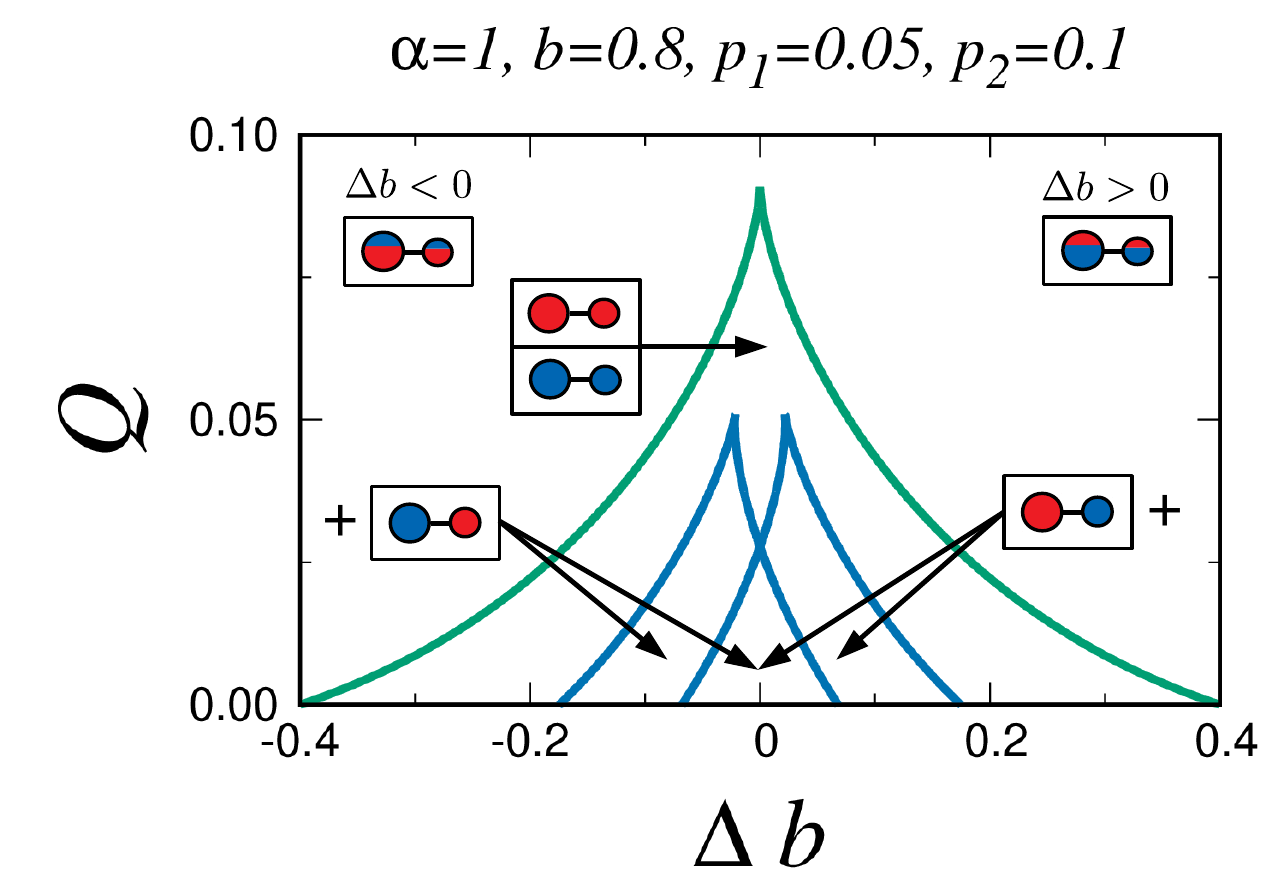}}

\subfloat[]{\label{fig:phases_pair:c}\includegraphics[width=0.9\textwidth]{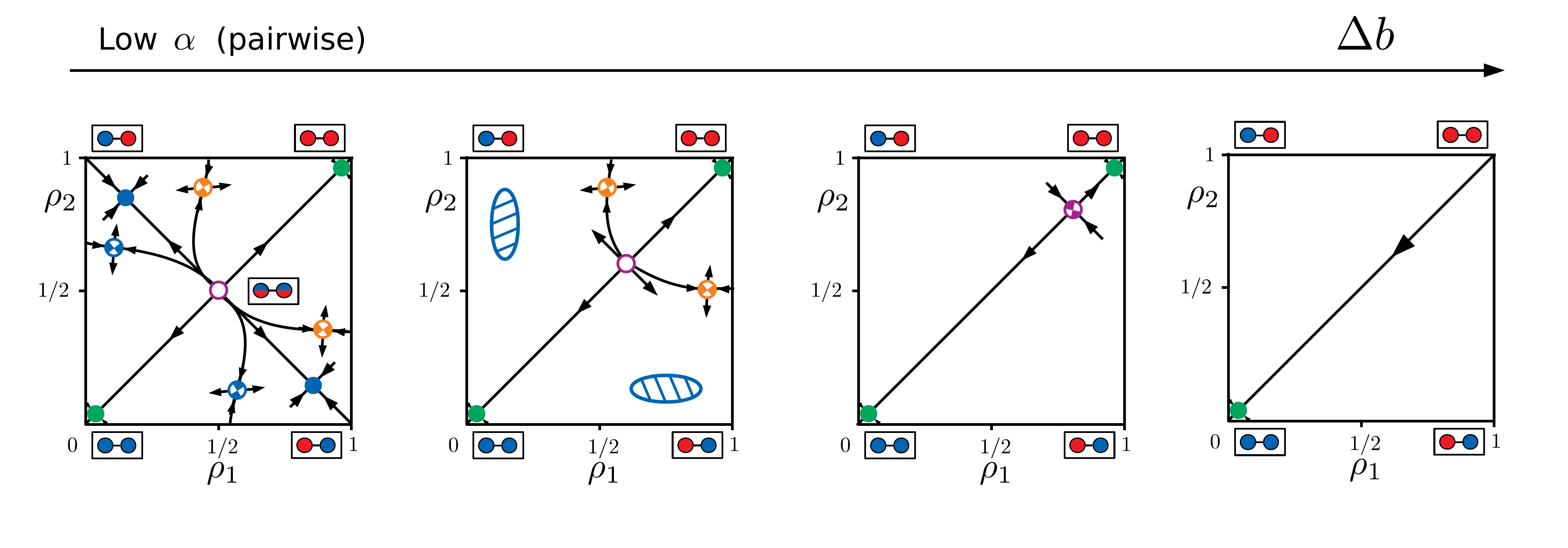}}
\caption{Phase diagrams for (a) $p_{1}=p_{2}=p=0.1$, and (b) $p_{1}=0.05$, $p_{2}=0.1$, and vector fields (c) for fixed values of the model parameters $\alpha=1$ (linear or pairwise regime) and $b=0.8$. In the phase diagrams (a,b) the varying parameters are $(Q, \Delta b)$, i.e. the noise and bias asymmetry. The transition lines (green and blue) delimit the parameter regions where the different possible fixed points are stable. Circles inside a square indicate the corresponding fixed point following the scheme of Fig. \ref{fig:color_scheme}. The phase diagram (a) corresponds to the case of two equal groups, while in (b) there is a majority (big circle) and minority (small circle) group. In the region below the green line both consensus states are stable, and above only one of them remains, while below the blue line polarization is stable.  In panel (c), the left vector field is a typical situation below the blue line of the phase diagram (a) for $\Delta b=0$. The other vector fields (from left to right) show how this changes as we increase $\Delta b$ and cross the various transition lines. The elliptical striped zones are regions where the dynamics is very slow and meta-stable states are possible (see Section \ref{sec_metastable}).}
\label{fig:phases_pair}
\end{figure}

In Fig. \ref{fig:phases_pair} we show the phase diagram and vectors fields in the prototypical scenario of pairwise interactions (low $\alpha$) as a function of bias asymmetry $\Delta b$. The phase diagrams in Fig. \ref{fig:phases_pair:a} and Fig. \ref{fig:phases_pair:b} correspond to the well-known \emph{cusp catastrophe} \cite{Stewart:1982}, where transitions are saddle node bifurcations in which two fixed points (stable and saddle point or saddle point and unstable) merge and disappear for high enough bias asymmetry. In the case of two equal groups ($p_{1}=p_{2}$, Fig. \ref{fig:phases_pair:a}), when tuning bias asymmetry, polarization is destroyed favoring the consensus states. After that, for a specific high value of the asymmetry, one of the two consensus states disappears (Fig. \ref{fig:phases_pair:c}) and the only remaining state is $\rho^{{\rm st}}_{1} = \rho^{{\rm st}}_{2} \approx 0$ for $\Delta b > 0$, or $\rho^{{\rm st}}_{1} = \rho^{{\rm st}}_{2} \approx 1$ for $\Delta b < 0$. Every time a pair of fixed points merge, there is a region in phase space where the dynamics $\rho_{1}(t)$, $\rho_{2}(t)$ becomes very slow and meta-stable states appear (in Section \ref{sec_metastable} we discuss this in more detail). Note that in the symmetric version of the model ($\Delta b=0$), when more than one stable fixed point exists, initial conditions determine the final state. However, when the exogenous ingredient of bias asymmetry is introduced by the social media ($\Delta b \neq 0$), it is possible to `select and control' the final opinion of the system.

Another relevant scenario is that of asymmetric groups $p_{1} \neq p_{2}$, where one of them is either better connected and/or bigger in size, i.e. for $p_{1}<p_{2}$ group $1$ has more nodes or links than group $2$, and the other way around for $p_{1}>p_{2}$. The two polarized states ($\rho^{{\rm st}}_{1} \approx 0$, $\rho^{{\rm st}}_{2} \approx 1$ and $\rho^{{\rm st}}_{1} \approx 1$, $\rho^{{\rm st}}_{2} \approx 0$) are not symmetric, depending on which is the opinion of the majority and minority groups. For this reason, there are two transition lines in the phase diagram of Fig. \ref{fig:phases_pair:b} with cusps at different positions, one for $\Delta b > 0$ and the other for $\Delta b < 0$. Thus, bias asymmetry promotes (instead of destroying) polarization in the region between the two cusps. This result has a clear interpretation: if the social platform favors the opinion of the minority group, polarization will become stronger as it will be harder for the majority group to convince the other, leading the system towards consensus. The value of $\Delta b$ at the cusp is the `optimal' one if we wish to balance such majority-minority scenario (i.e., asymmetry in group sizes and connections) by using an exogenous algorithmic bias. 



\begin{figure}[ht]
\centering
\includegraphics[width=0.6\textwidth]{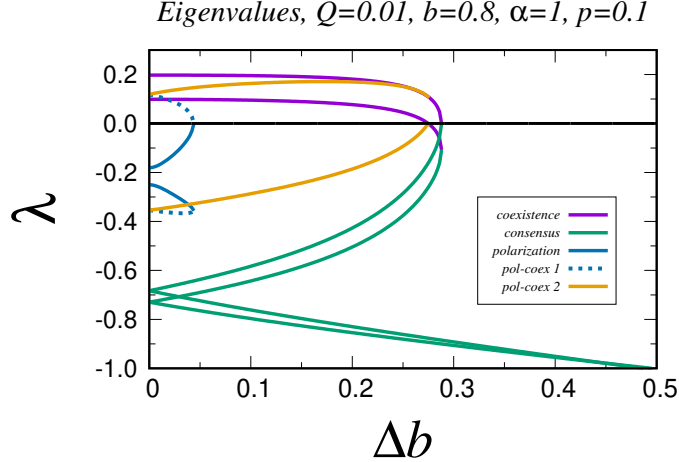}
\caption{Eigenvalues of the various fixed points of Fig. \ref{fig:color_scheme} and Fig. \ref{fig:phases_pair:c} for parameter values $\alpha=1$ (linear or pairwise regime), $b=0.8$, $p_{1}=p_{2}=p=0.1$ and $Q=0.01$, as a function of the bias asymmetry $\Delta b$. The color and name coding in the legends is equivalent to that of Fig. \ref{fig:color_scheme}. The lines of the same color (blue) correspond to a pair of polarized fixed points, dashed (saddle point) and solid (stable), that merge together and disappear for a particular value of $\Delta b$.}
\label{fig:eigenvalues_pair}
\end{figure}

In Fig. \ref{fig:eigenvalues_pair} we plot the eigenvalues of all fixed points as a function of bias asymmetry in the same case specified in Fig. \ref{fig:phases_pair}. The eigenvalues provide us with a lot of information about the nature of the fixed points and the dynamics close to them. The sign of the eigenvalues in Fig. \ref{fig:eigenvalues_pair} agree with the schematic representation of the vector fields in Fig. \ref{fig:phases_pair:c}, and it determines the stability analysis of the fixed points. Every time an eigenvalue becomes zero, a pair of fixed points disappears, defining a transition or bifurcation.

\begin{figure}[ht]
\centering
\subfloat[]{\label{fig:phases_group:a}\includegraphics[width=0.47\textwidth]{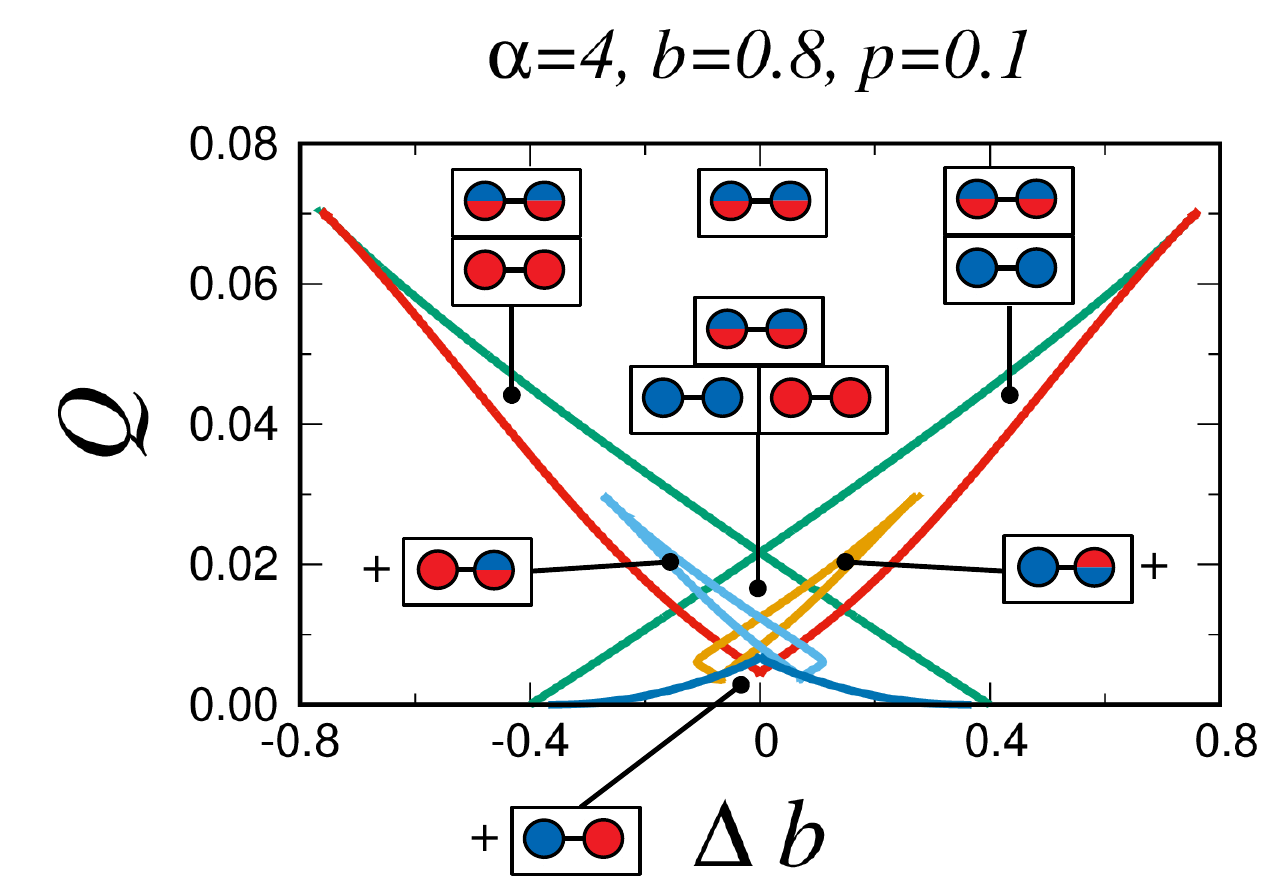}}
\subfloat[]{\label{fig:phases_group:b}\includegraphics[width=0.47\textwidth]{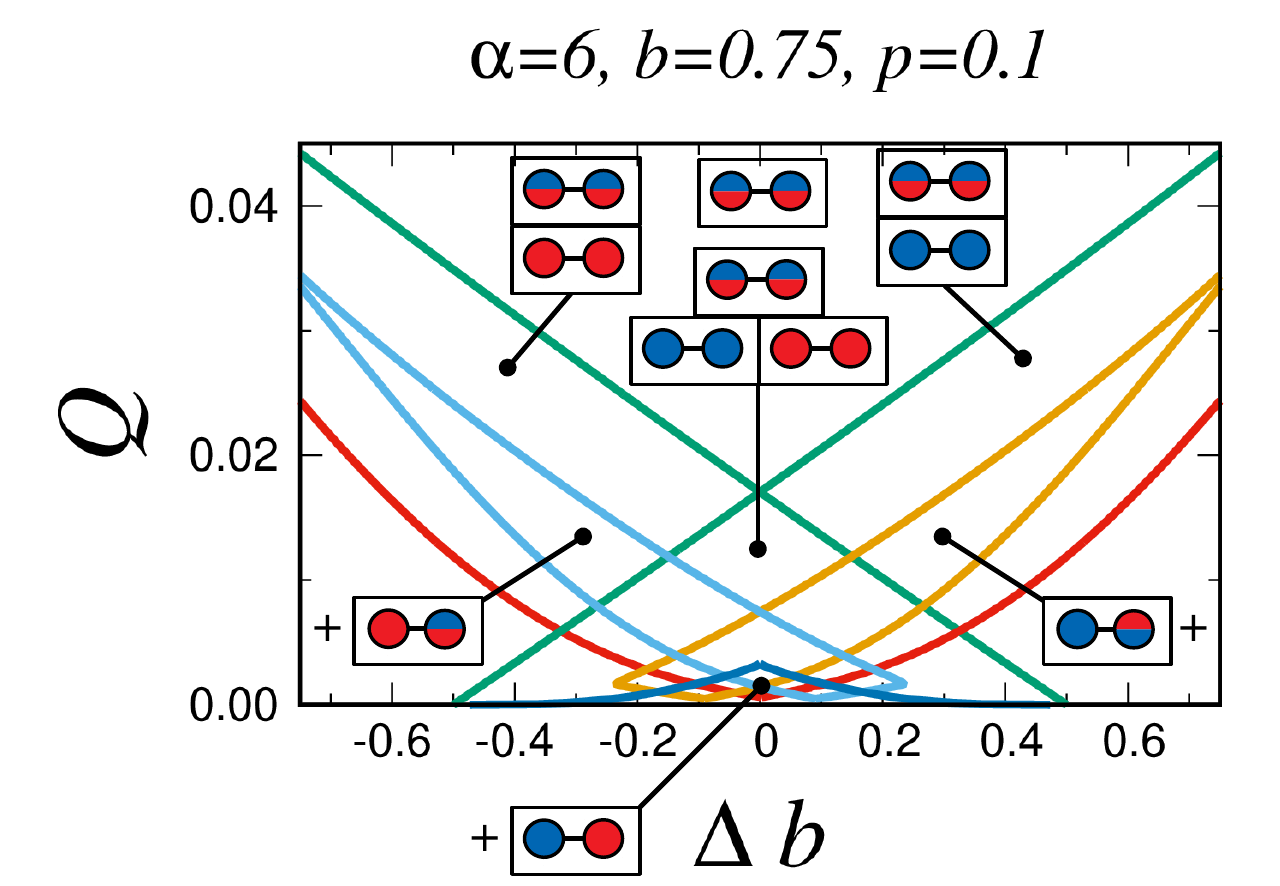}}

\subfloat[]{\label{fig:phases_group:c}\includegraphics[width=0.90\textwidth]{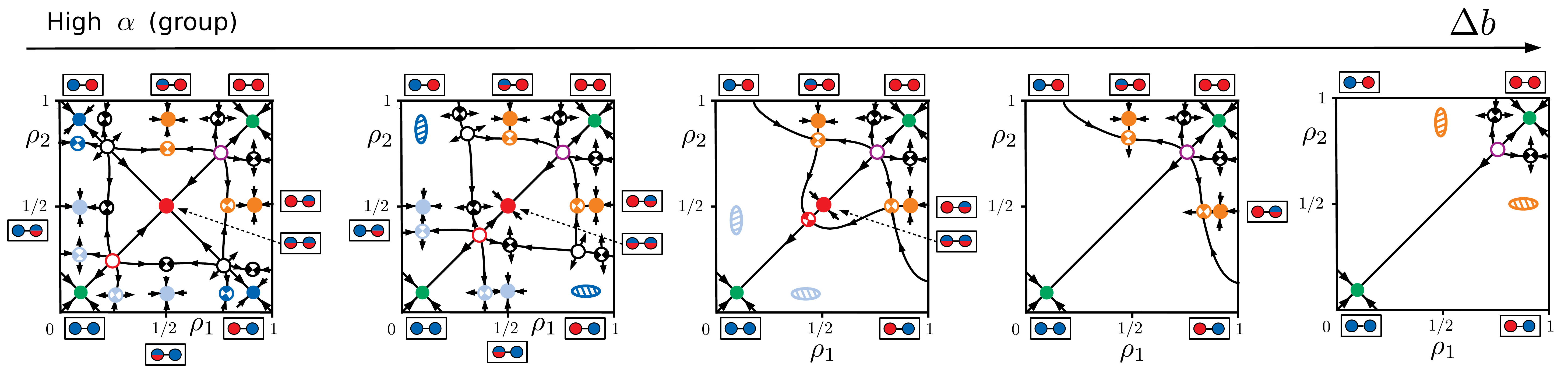}}
\caption{Phase diagrams for (a) $\alpha=4$, $b= 0.8$, and (b) $\alpha=6$, $b= 0.75$, and vector fields (c) for fixed connectivity $p_{1}=p_{2}=p=0.1$. In the phase diagrams (a,b) the varying parameters are $(Q, \Delta b)$, i.e., the noise and bias asymmetry. The transition lines (green, red, dark and light blue, and yellow) delimit the parameter regions where fixed points are stable. Circles inside a square indicate the corresponding fixed points following the scheme of Fig. \ref{fig:color_scheme}. In panel (c), the left vector field is a typical situation below the blue line (and above the red, light blue and yellow lines) of the phase diagram (b) for $\Delta b=0$. The other vector fields from left to right show how this changes as we increase $\Delta b$ and cross the various transition lines. The elliptical striped zones are regions where the dynamics is very slow and meta-stable states are possible (see Section \ref{sec_metastable}).}
\label{fig:phases_group}
\end{figure}

In Fig. \ref{fig:phases_group} we show the phase diagram and vector field in the case of group interactions (high $\alpha$) as a function of bias asymmetry $\Delta b$. Figs. \ref{fig:phases_group:a} and \ref{fig:phases_group:b} correspond to the so-called \emph{butterfly catastrophe} \cite{Stewart:1982}. We observe some differences with respect to the pair interaction case, besides the rich phenomenology of additional fixed points. The first difference is that the coexistence and consensus states can be both stable for some parameter values, at odds with results in Fig. \ref{fig:phases_pair}. With respect to the dependence of the consensus states on bias asymmetry, for low noise we have a similar behavior as for pair interactions, while for high noise it is possible that a consensus state, which is not observed for $\Delta b=0$, appears for some value $\Delta b \neq 0$. Standard polarization is also destroyed for a critical value of the asymmetry in the group interaction case. The difference is that new stable polarized states are possible in the group case (pol-coex 2 and pol-coex 4 in Fig. \ref{fig:color_scheme}), whose behavior with respect to bias asymmetry is non-trivial. Similarly to the consensus states, these new polarized states may appear for a particular value of the asymmetry, even though they are not present in the symmetric case.

\begin{figure}[ht]
\centering
\subfloat[]{\label{fig:eigenvalues_group:a}\includegraphics[width=0.47\textwidth]{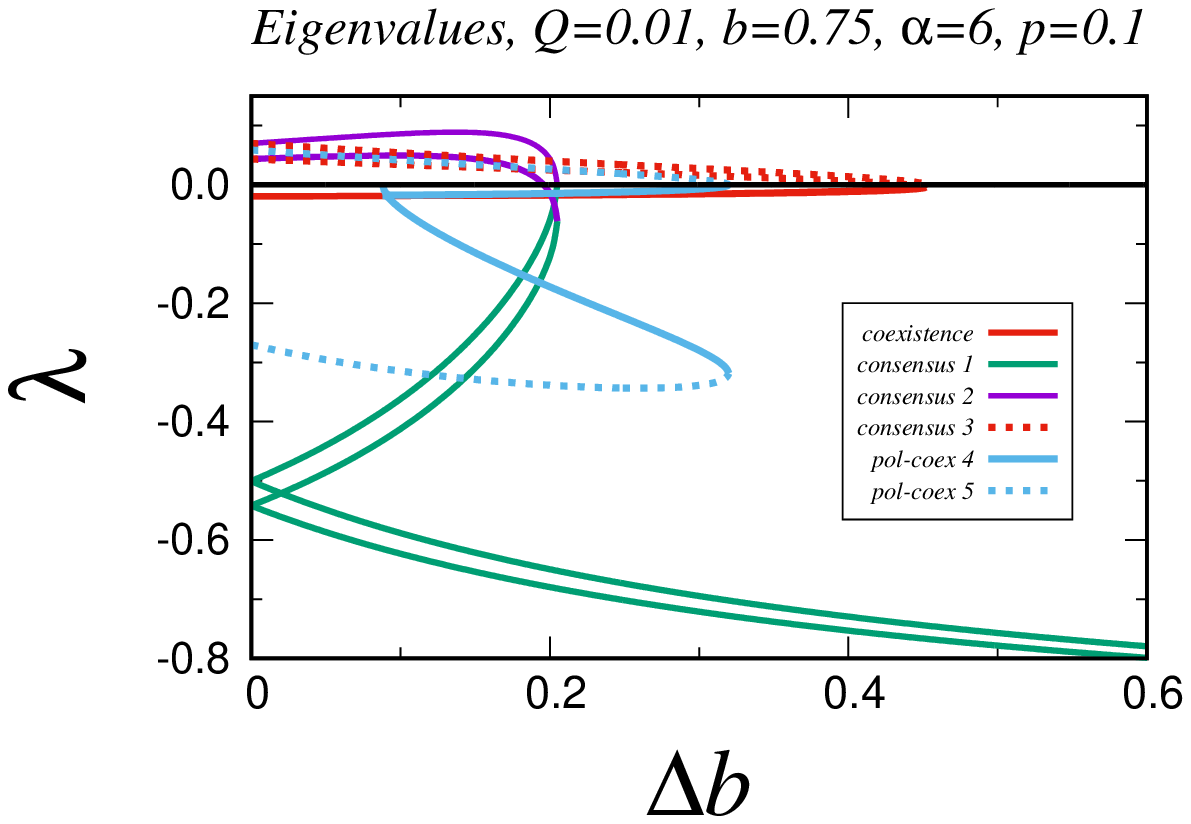}}
\subfloat[]{\label{fig:eigenvalues_group:b}\includegraphics[width=0.47\textwidth]{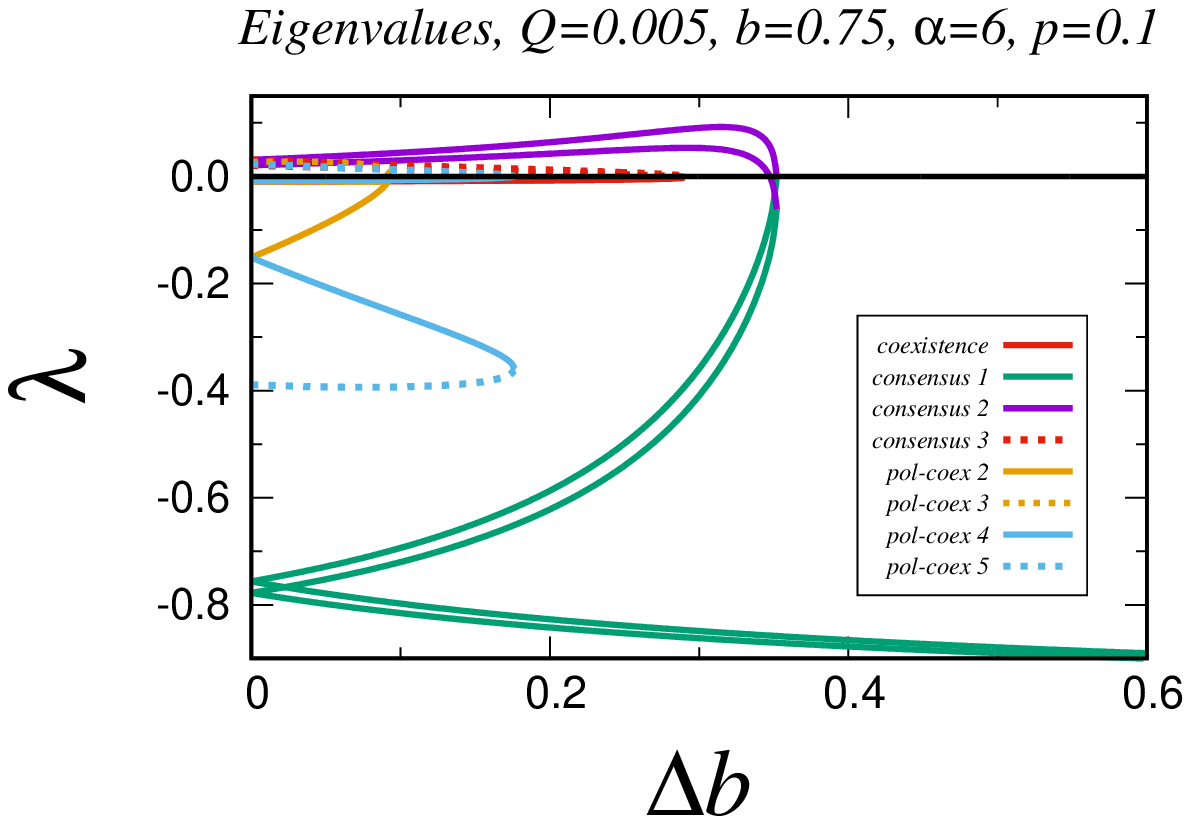}}

\subfloat[]{\label{fig:eigenvalues_group:c}\includegraphics[width=0.47\textwidth]{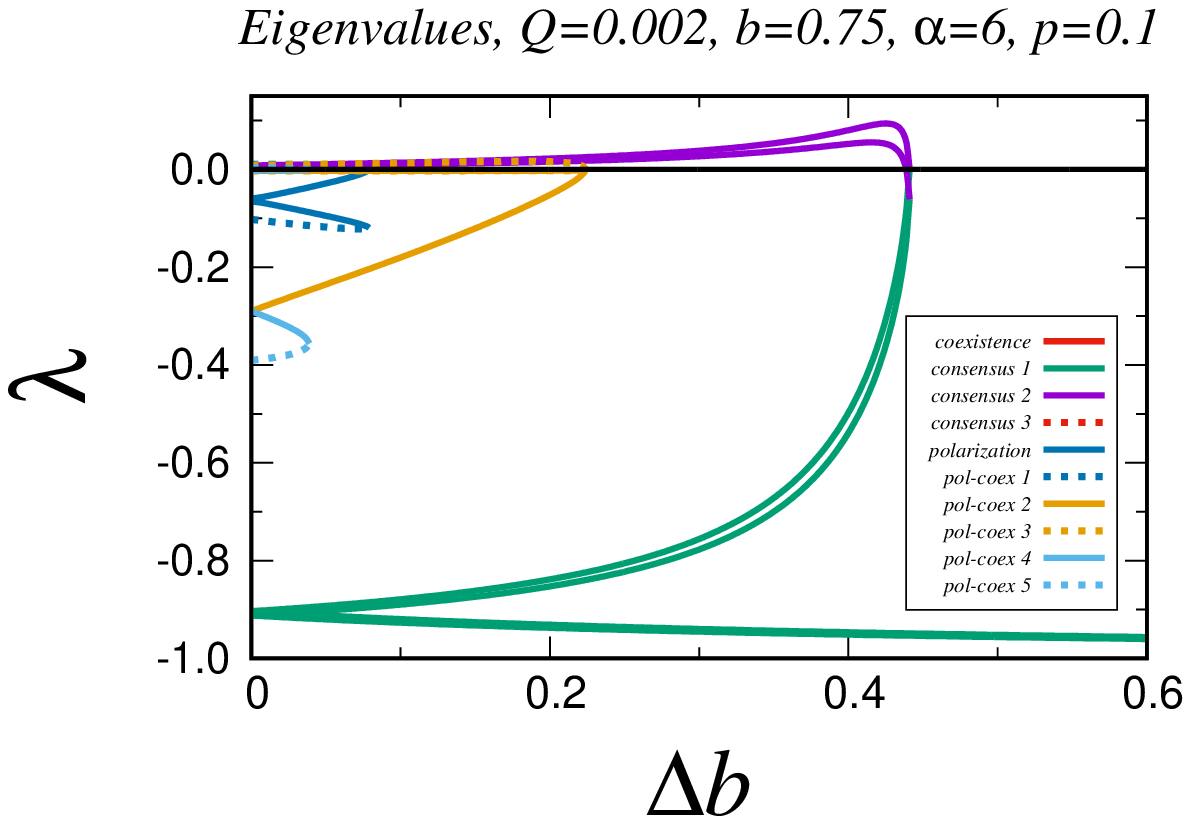}}
\caption{Eigenvalues of the various fixed points of Fig. \ref{fig:color_scheme} and Fig. \ref{fig:phases_group:c} for parameter values $\alpha=6$ (group regime), $b=0.75$, $p_{1}=p_{2}=p=0.1$ and $Q=0.01, 0.005, 0.002$, as a function of bias asymmetry $\Delta b$. The color and name coding in the legends is equivalent to that of Fig. \ref{fig:color_scheme}. Lines of the same color (light and dark blue, red and yellow) correspond to a pair of fixed points, dashed (saddle point or unstable) and solid (stable), that merge together and disappear for a particular value of $\Delta b$.}
\label{fig:eigenvalues_group}
\end{figure}

In Fig. \ref{fig:eigenvalues_group} we show the eigenvalues of the fixed points of Fig. \ref{fig:color_scheme} and Fig. \ref{fig:phases_group:c} as a function of $\Delta b$ for different values of the noise $Q$. These figures provide us with information about the stability, dynamics and transitions that are possible in this case. Note that, depending on the value of $Q$, we may find some of the fixed points or not, and that different transitions happen for specific values of $\Delta b$, in accordance with the phase diagram in Fig. \ref{fig:phases_group:b}. The vector field scheme in Fig. \ref{fig:phases_group:c} is a low-noise scenario where all fixed points are present (Fig. \ref{fig:eigenvalues_group:c}). For other values of the noise $Q$ (in Fig. \ref{fig:eigenvalues_group:a} and Fig. \ref{fig:eigenvalues_group:b}), not all the fixed of Fig. \ref{fig:phases_group:c} are possible, and the transitions may occur in different orders as we increase $\Delta b$.

\section{Global dynamics, convergence times and meta-stable states}\label{sec_metastable}

The global dynamics $\rho_{1}(t)$, $\rho_{2}(t)$ of the system has a non-trivial dependence on the initial condition $\rho_{1}(0)$, $\rho_{2}(0)$ and the model parameters $(Q, \alpha, b, \Delta b, p_{1}, p_{2})$, besides time $t$. The determination of the fixed points, the stability, and the local (linearized) dynamics are a good guideline to predict and understand, at least qualitatively, the dynamical behavior of the system. There are other aspects of the dynamics that cannot be fully explained by the fixed points and the linear dynamics approach. Among these, we study with particular attention what we call {\it meta-stable states}, where the dynamics slows down strongly and stays for a long time around a determined value of the state variables. This phenomenon is observed in the model, especially when two fixed points merge and disappear (the elliptical striped zones in Fig. \ref{fig:phases_pair:c} and Fig. \ref{fig:phases_group:c}). As there are no fixed points around these zones, it is not possible to evaluate the eigenvalues and explore the local dynamics. Thus, we need to use a different theoretical method to characterize the meta-stable states. We also explore the time needed to reach the final states and the dependence on the initial conditions, theoretically and by means of Monte Carlo simulations.

\subsection{Numerical simulations}\label{sec_num}

\begin{figure}[ht]
\centering

\subfloat[]{\label{fig:Traj_hom:a}\includegraphics[width=0.32\textwidth]{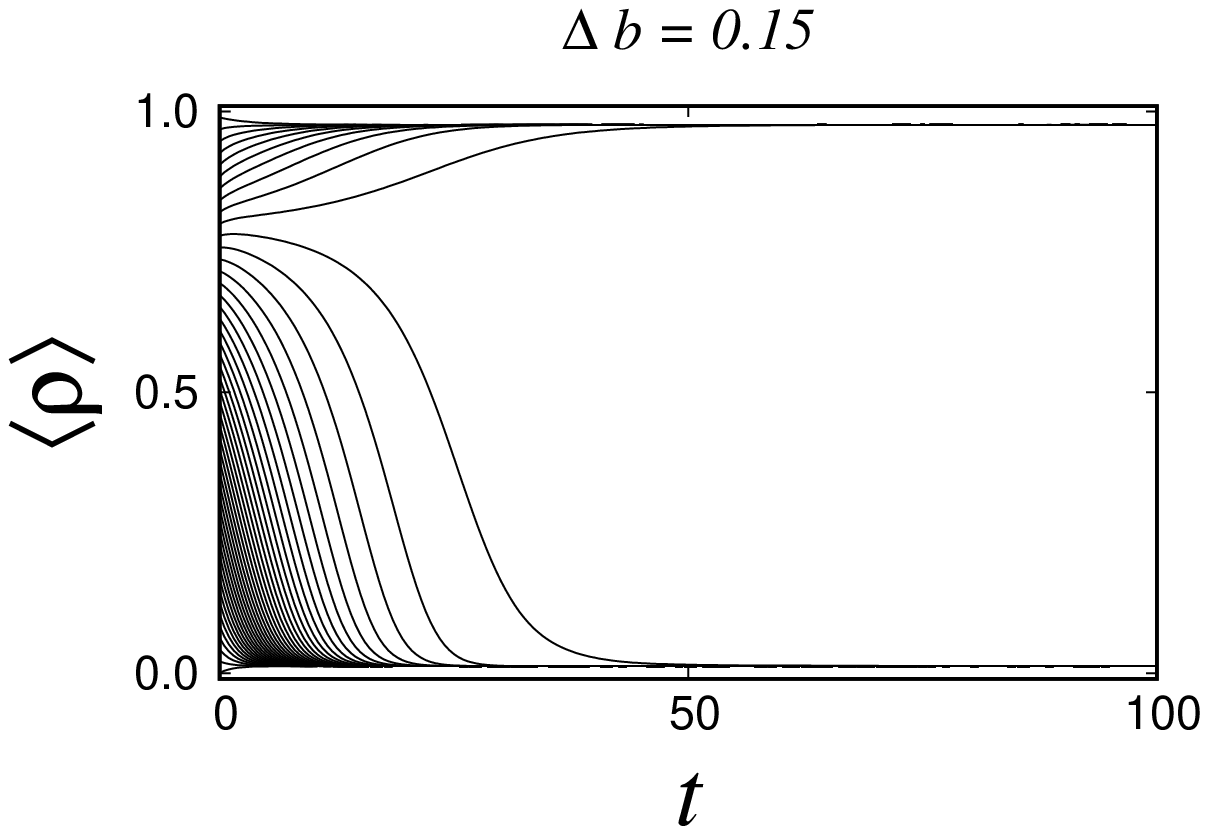}}
\subfloat[]{\label{fig:Traj_hom:b}\includegraphics[width=0.32\textwidth]{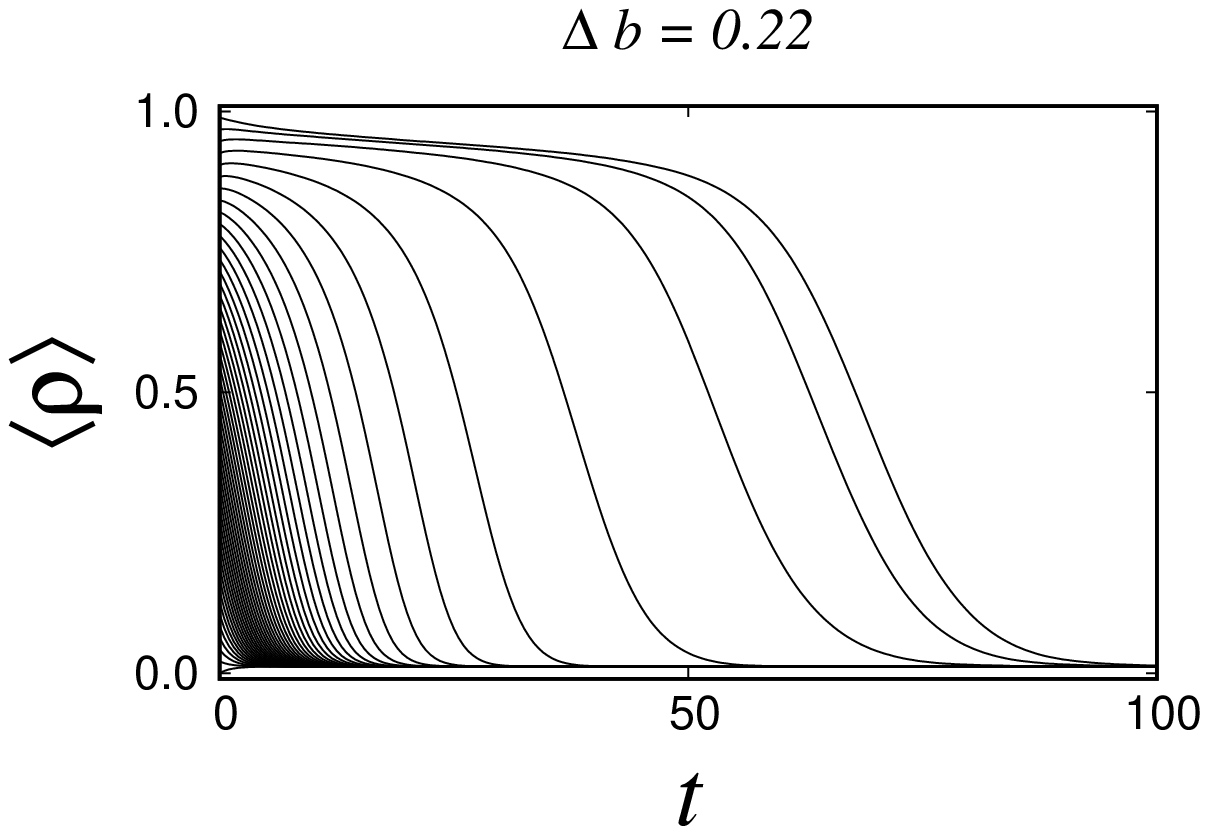}}
\subfloat[]{\label{fig:Traj_hom:c}\includegraphics[width=0.32\textwidth]{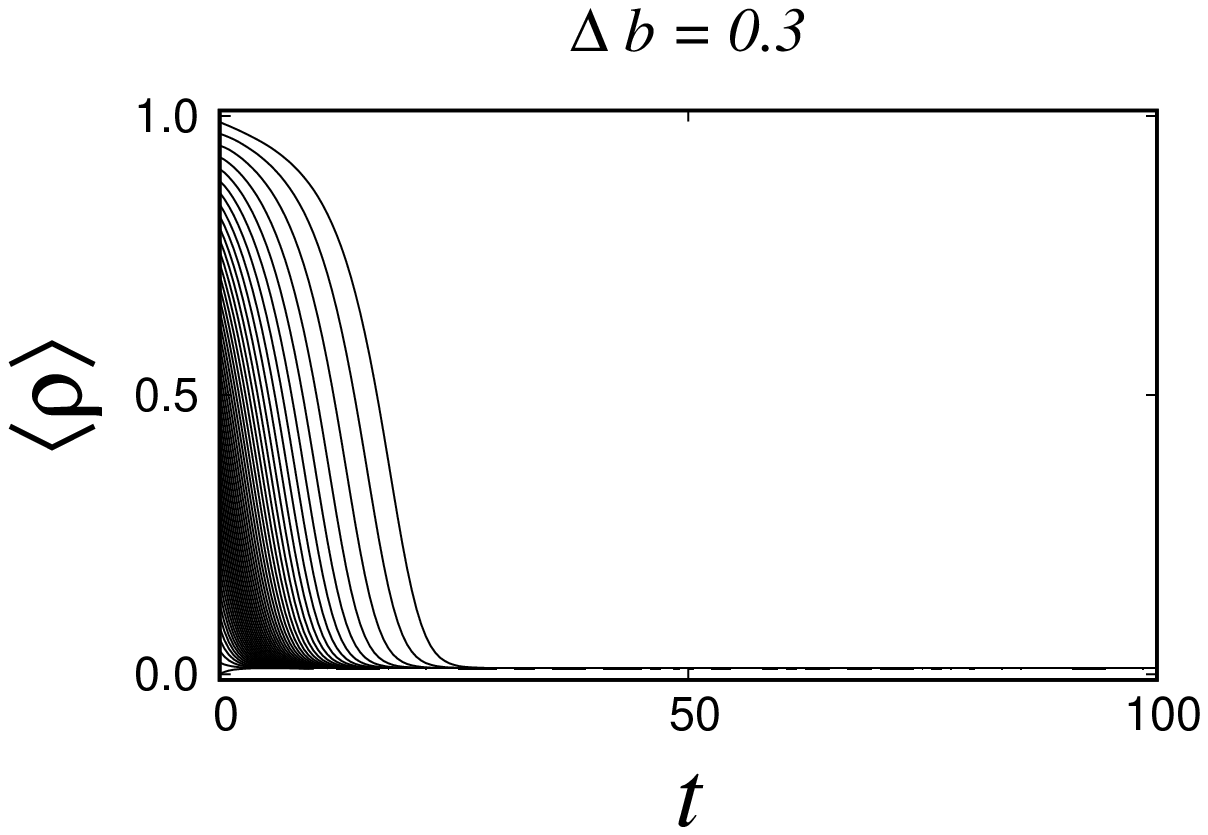}}

\subfloat[]{\label{fig:Traj_hom:d}\includegraphics[width=0.32\textwidth]{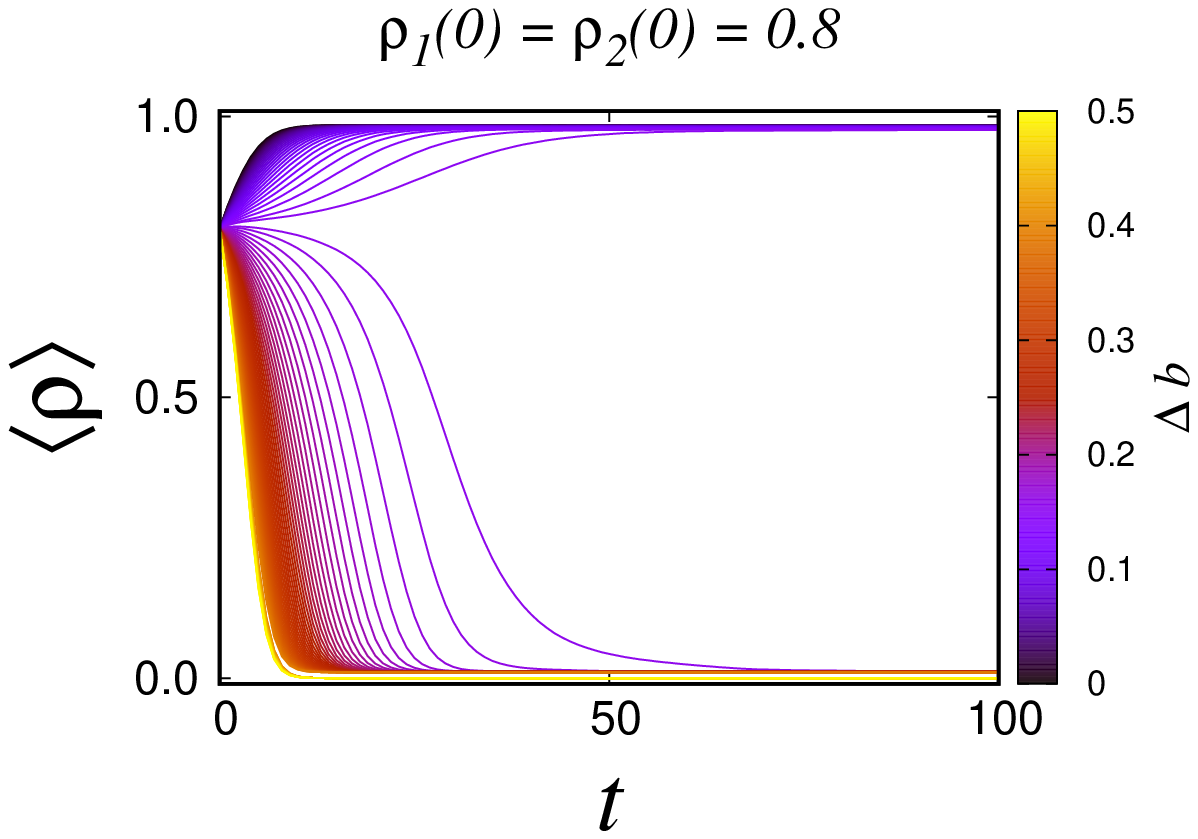}}
\subfloat[]{\label{fig:Traj_hom:e}\includegraphics[width=0.32\textwidth]{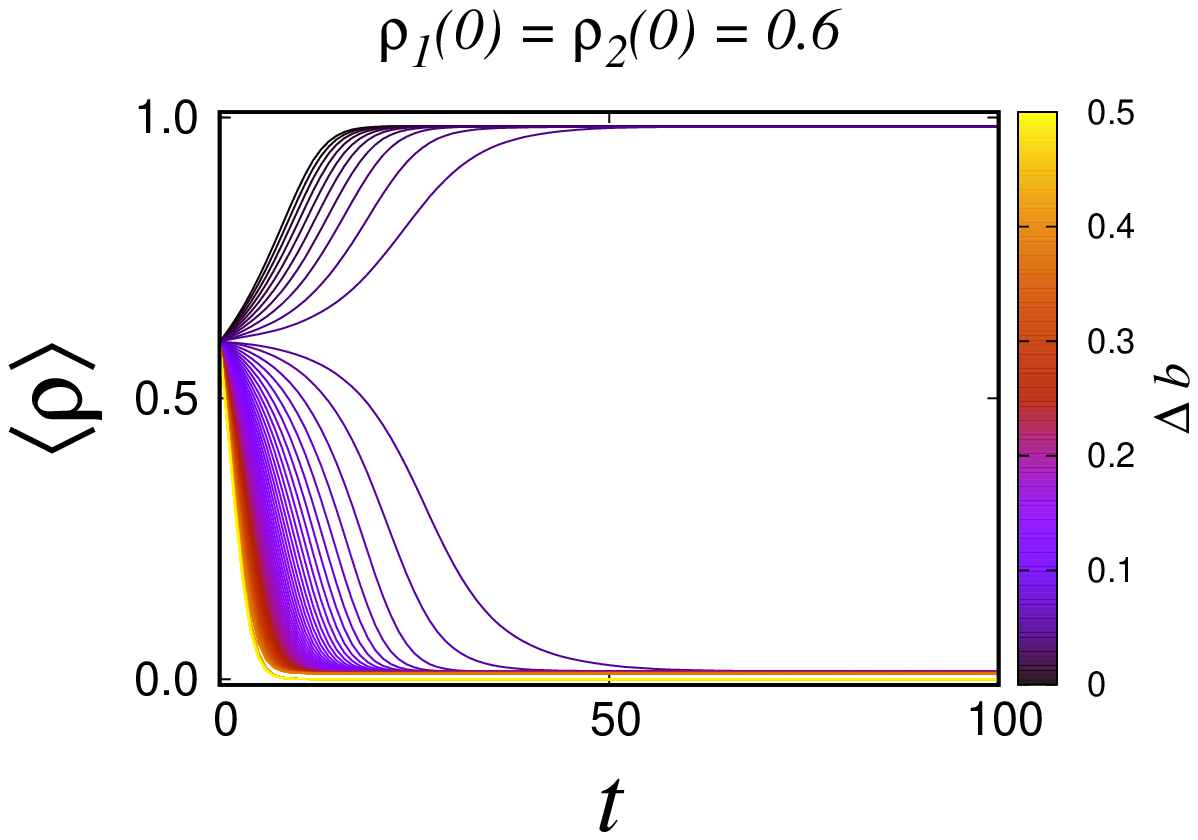}}
\subfloat[]{\label{fig:Traj_hom:f}\includegraphics[width=0.32\textwidth]{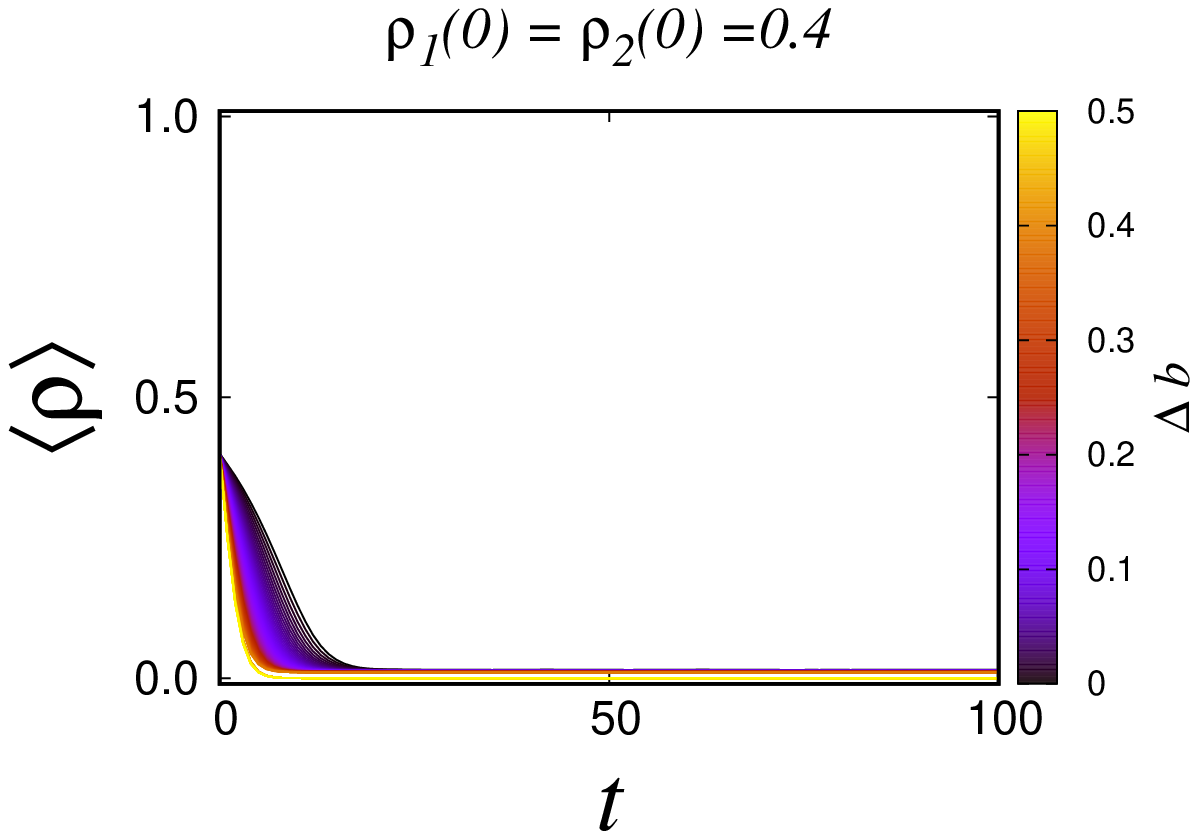}}

\caption{Average density $\langle \rho \rangle$ of nodes in state $1$ as a function of time $t$, starting from a homogeneous initial condition $\rho_{1}(0) = \rho_{2}(0) = \rho(0)$. The model and connectivity parameters are $\alpha=1$, $b=0.8$, $Q=0.01$, $p_{1}=p_{2}=0.1$, $z_{1}=z_{2}=18$, $z_{12}=z_{21}=2$, and the size of the network is $N=20000$, divided in two groups (communities) of equal size ($N_{1}=N_{2}=N/2$), averaged over $1000$ realizations of the dynamics. In the top panels (a-c), trajectories correspond to various initial conditions $\rho(0)$ for a fixed value of the bias asymmetry $\Delta b$ (specified in the title). In the bottom panels (d-f), a color gradient is used for bias asymmetry $\Delta b$ in the range $[0, 0.5]$ for a fixed initial condition $\rho_{1}(0)$, $\rho_{2}(0)$ (specified in the title).}
\label{fig:Trajectories_hom}
\end{figure}

\begin{figure}[ht]
\centering

\subfloat[]{\label{fig:Traj_pol:a}\includegraphics[width=0.32\textwidth]{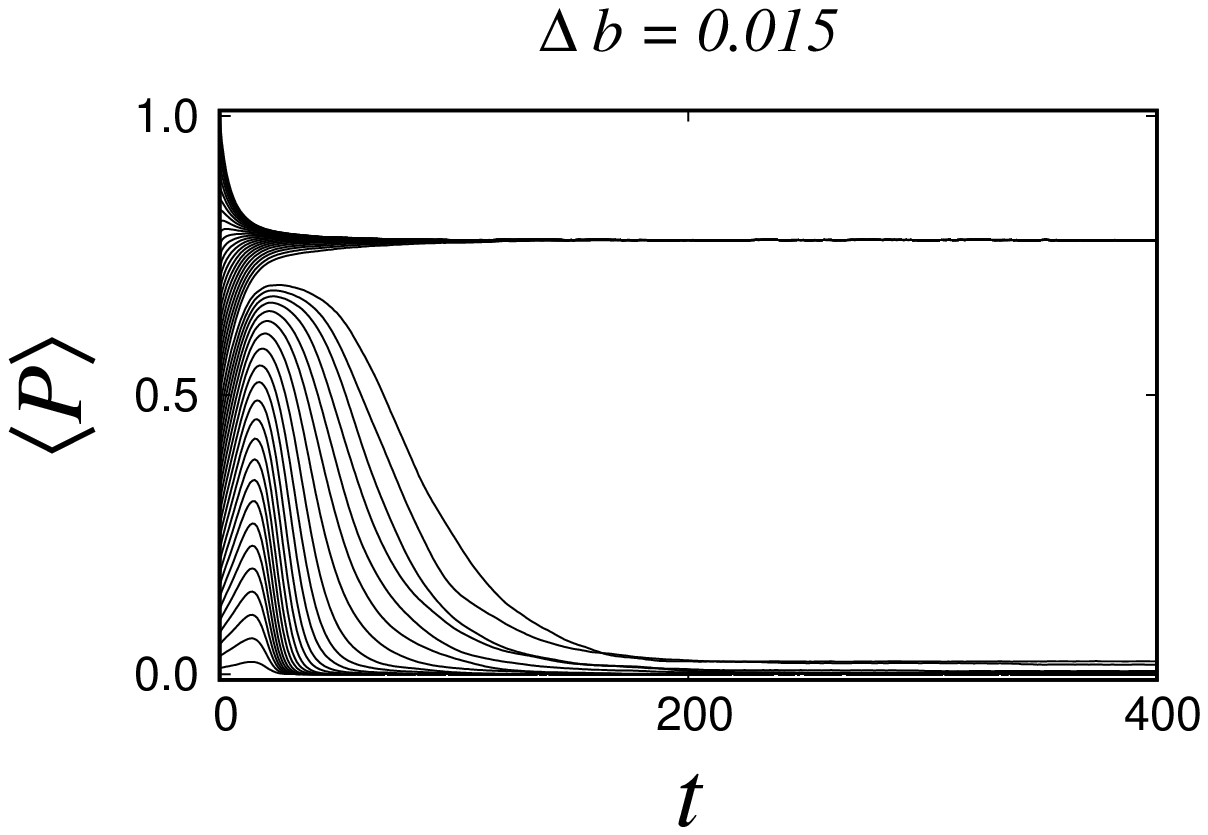}}
\subfloat[]{\label{fig:Traj_pol:b}\includegraphics[width=0.32\textwidth]{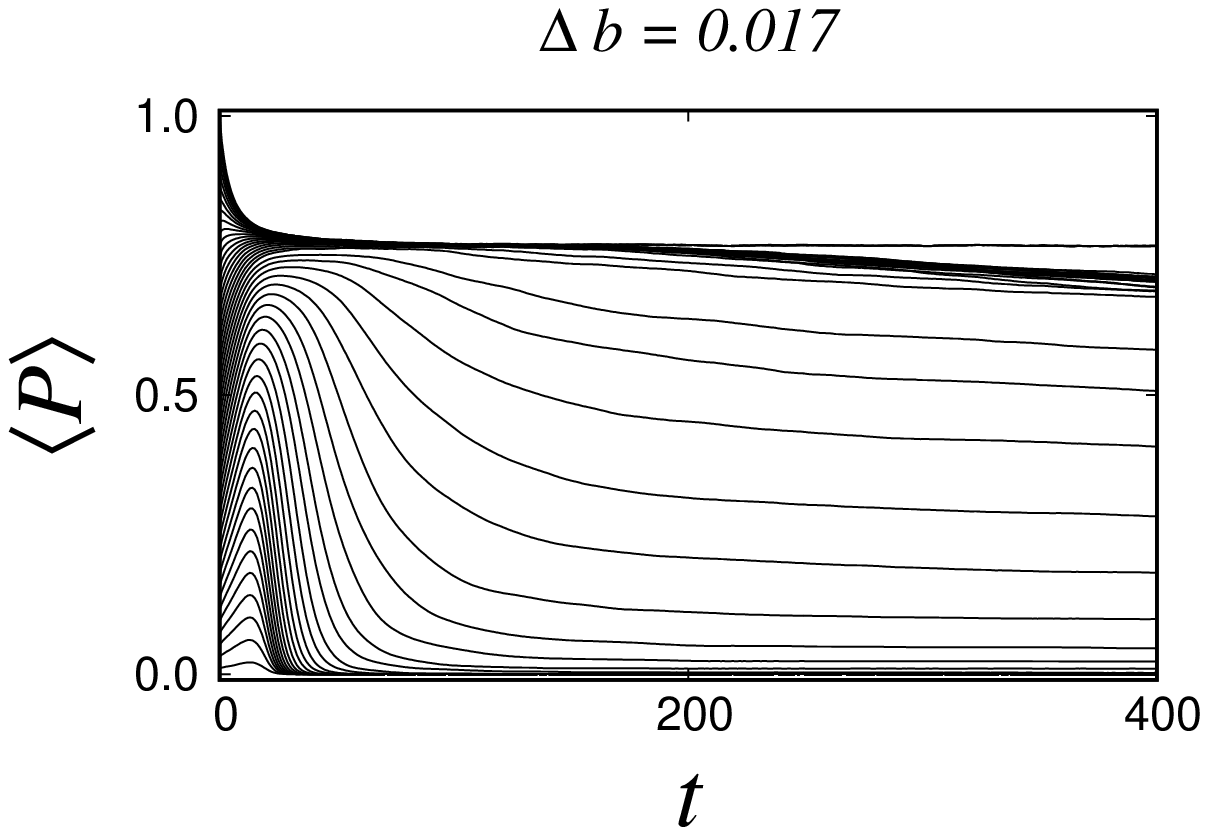}}
\subfloat[]{\label{fig:Traj_pol:c}\includegraphics[width=0.32\textwidth]{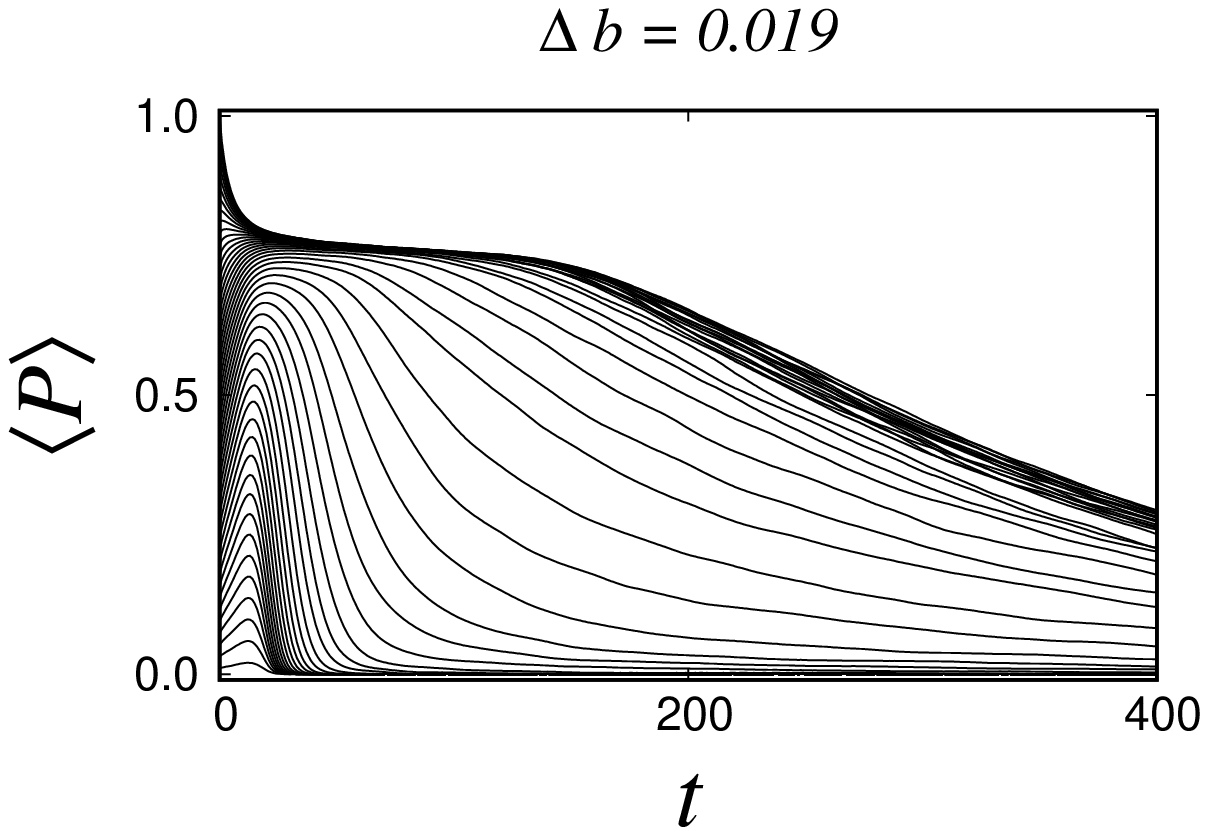}}

\subfloat[]{\label{fig:Traj_pol:d}\includegraphics[width=0.32\textwidth]{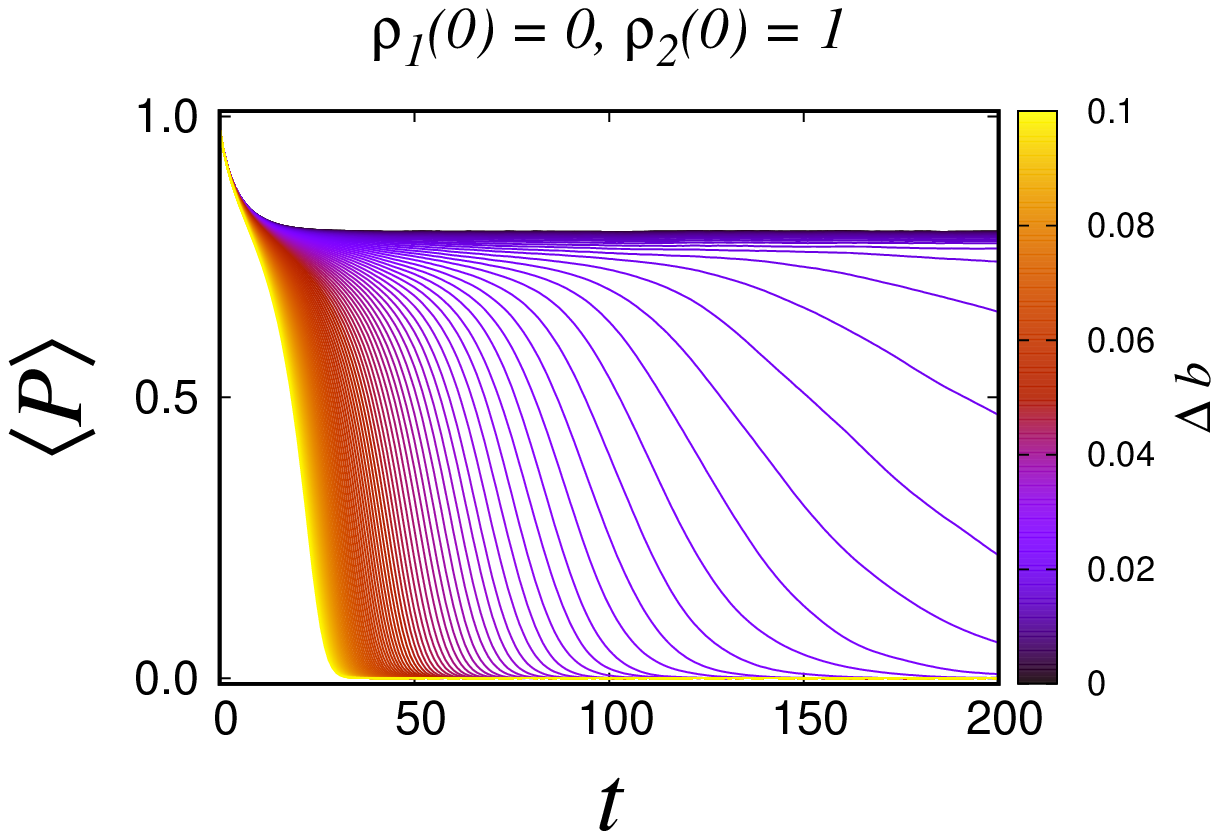}}
\subfloat[]{\label{fig:Traj_pol:e}\includegraphics[width=0.32\textwidth]{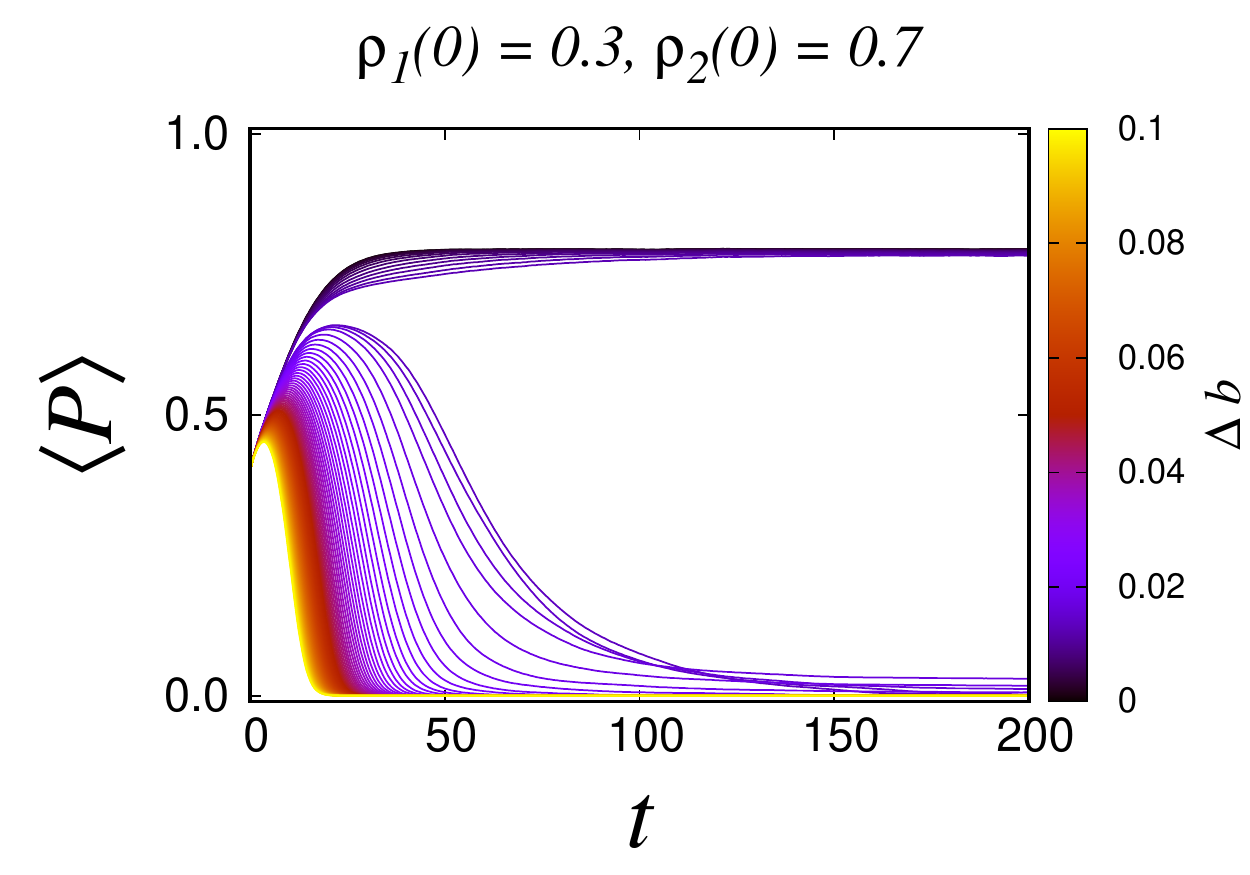}}
\subfloat[]{\label{fig:Traj_pol:f}\includegraphics[width=0.32\textwidth]{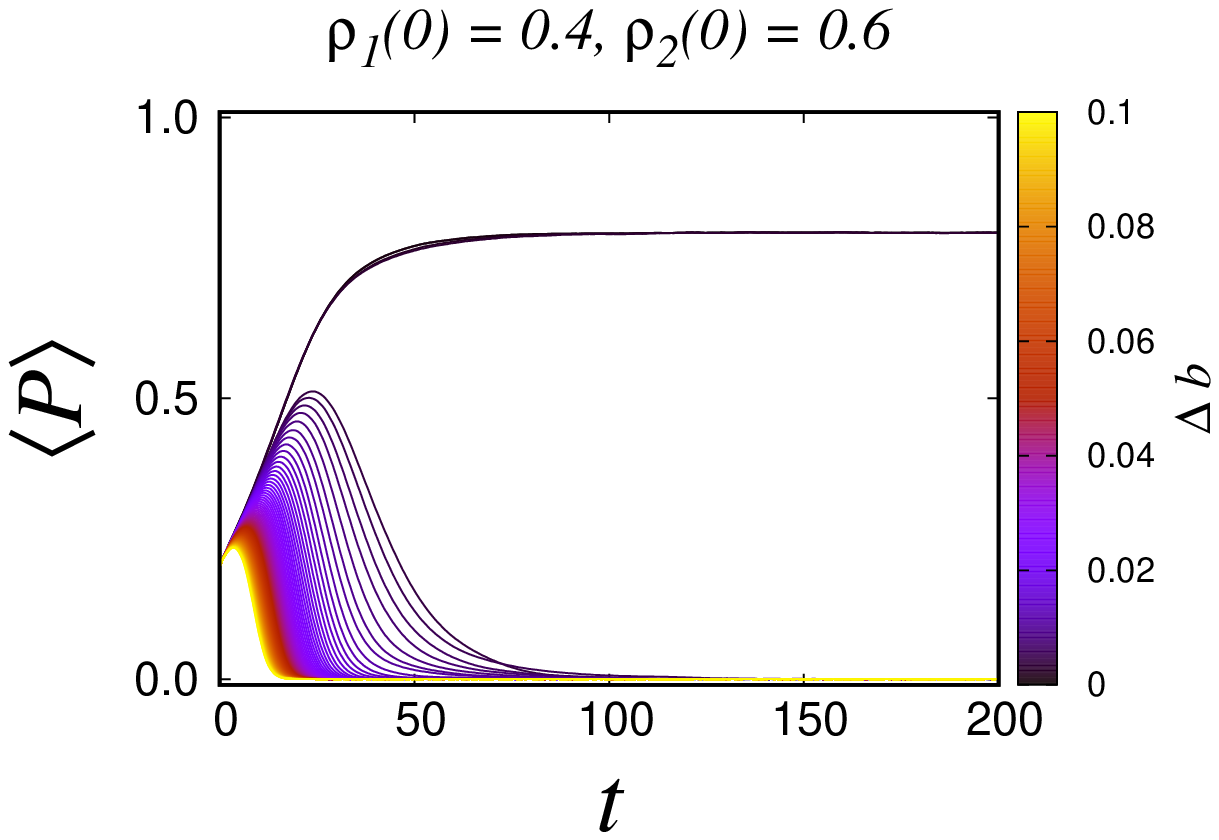}}

\caption{Average polarization $\langle P \rangle =\langle \vert \rho_{1} - \rho_{2}\vert \rangle$, starting from a polarized initial condition $\rho_{2}(0) = 1-\rho_{1}(0)$. All model, connectivity and network parameters are the same as in Fig. \ref{fig:Trajectories_hom}. In the top panels (a-c) a fixed bias asymmetry is used for different initial conditions, while in the bottom panels (d-f) a fixed initial condition is used for different bias asymmetries (colors).}
\label{fig:Trajectories_pol}
\end{figure}

Before introducing the theoretical description of the meta-stable states, we analyze the results coming from Monte Carlo simulations. Implementing the rules of the model (Section \ref{sec_model_def}), we obtain stochastic trajectories $\rho_{1}(t)$, $\rho_{2}(t)$, from which we calculate the average global state of the system $\langle \rho(t) \rangle$ and polarization $\langle P(t) \rangle$ that characterize the dynamics. In Fig. \ref{fig:Trajectories_hom} and Fig. \ref{fig:Trajectories_pol} we show numerical results for pair interactions ($\alpha=1$) on a large ($N=20000$) $z-$regular type of network with modular structure. In order to compare the simulations with the phenomenology coming from the theory, we use the same parameter values as in Fig. \ref{fig:phases_pair:a} and Fig. \ref{fig:eigenvalues_pair}, i.e. $Q=0.01$, $b=0.8$ and $p_{1}=p_{2}=0.1$. 

In Fig. \ref{fig:Trajectories_hom} we show the dynamics for various homogeneous initial conditions and bias asymmetries. In the top panels, from left to right, one of the consensus states becomes unstable for a determined value of the bias asymmetry and then, independently on the initial condition, all trajectories evolve towards the remaining stable consensus state. This corresponds to crossing (horizontally) the green transition line in the phase diagram of Fig. \ref{fig:phases_pair:a}. Note that before the transition, when the two consensus states are possible, and depending on the initial condition $\rho_{1}(0)=\rho_{2}(0)=\rho(0)$, the dynamics evolves towards one state or the other. We can thus define a threshold initial condition $\rho_{0}$ that separates the basin of attraction of the consensus states. This threshold depends on the bias asymmetry $\rho_{0}(\Delta b)$: for no asymmetry ($\Delta b=0$) it is $\rho_{0}=0.5$, it increases for $\Delta b>0$ (decreases for $\Delta b<0$), and it is not defined above the transition point as only one consensus state is stable. In Fig. \ref{fig:Trajectories_pol}, we show the dynamics for different polarized initial conditions and bias asymmetries. In the top panels, from left to right, the polarized state becomes unstable for a determined value of the bias asymmetry and then all trajectories evolve towards a non-polarized ($P=0$) consensus state. This corresponds to crossing (horizontally) the blue transition line in the phase diagram of Fig. \ref{fig:phases_pair:a}. Note that the dynamical results shown in Fig. \ref{fig:Trajectories_hom} and Fig. \ref{fig:Trajectories_pol} are in good agreement with the qualitative description of the vector fields in Fig. \ref{fig:phases_pair:c}. 

A significant dynamic phenomenon observed in Figs. \ref{fig:Trajectories_hom} and \ref{fig:Trajectories_pol} (specially in panels \ref{fig:Traj_hom:b}, \ref{fig:Traj_pol:b}, \ref{fig:Traj_pol:c} and \ref{fig:Traj_pol:d}) is the presence of trajectories that get trapped for a long period of time in some state, but eventually get released and end in one of the possible final (stable) states. In Fig. \ref{fig:Traj_pol:d} we see that such a meta-stable state appears above the transition point $\Delta b^{*}$, and that its duration decreases as we increase the asymmetry. We represent this meta-stable state as an elliptical stripped zone in the vector fields Fig. \ref{fig:phases_pair:c}, and we characterize them theoretically in what follows (Section \ref{sec_meta}).

\subsection{Saddle node bifurcations and meta-stability}\label{sec_meta}

Close to a critical (bifurcation) point, it is possible to obtain the normal form of the dynamics \cite{Guckenheimer:2002}, which goes beyond the linearization of Eqs. (\ref{rho1_lin}, \ref{rho2_lin}). Assume that we have a fixed point $\rho^{{\rm st}}_{1}$ and $\rho^{{\rm st}}_{2}$ with an eigenvalue equal to zero $\lambda_{1}=0$, for a determined value of a tuning parameter, e.g. $\Delta b = \Delta b^{*}$ (bifurcation or critical point). We perform a change of variables to the eigenvector basis $\vec{u}(t) = P^{-1} \vec{\rho}(t)$, where the columns of the matrix $P$ are the eigenvectors $\vec{v}_{1,2}$, that is,

\begin{equation}
P=\left( \begin{array}{cc}
v_{11} & v_{21} \\
v_{12} & v_{22} 
\end{array} \right).
\end{equation}
The evolution of the transformed variables $\vec{u}(t)$ becomes
\begin{equation}
\label{general_dyn}
\frac{d \vec{\rho}}{d t} = \vec{\mu}(\Delta b, \vec{\rho}) \hspace{0.2cm} \rightarrow \hspace{0.2cm} \frac{d \vec{u}}{d t} = \vec{U}(\Delta b,\vec{u}) \equiv P^{-1} \vec{\mu}(\Delta b, P\vec{u}).
\end{equation}
The transformed vector field fulfills $\vec{U}(\Delta b^{*},\vec{u}_{{\rm st}})=0$ (here $\vec{u}_{{\rm st}}$ refers only to the fixed point at the bifurcation point $\Delta b=\Delta b^{*}$) and $\partial U_{i}/\partial u_{j} = -\lambda_{i} \delta_{ij}$ (again at $\Delta b^{*}$ and $\vec{u}_{{\rm st}}$), i.e., the linear part is uncoupled. In this case, when $\lambda_{1}=0$, there is a center manifold $\vec{u}(t)=\vec{h}(\Delta b, u_{1}(t))$, i.e. a special trajectory where the time dependence of all variables is governed by the slow $u_{1}(t)$. This satisfies $\vec{u}_{{\rm st}}=\vec{h}(\Delta b^{*},u^{{\rm st}}_{1})$ and the orthogonality condition $\partial h_{i}/\partial u_{1} = 0$. The center manifold can be obtained from Eq. (\ref{general_dyn}) as a series expansion. Once the functions $\vec{h}(\Delta b, u_{1})$ have been determined, we obtain a single equation for $u_{1}(t)$,
\begin{eqnarray}
\label{normal_form}
\frac{d u_{1}}{d t} &=& U_{1}(\Delta b, \vec{h}(\Delta b, u_{1})) \nonumber\\
&=& \beta^{(10)} (\Delta b-\Delta b^{*}) + \beta^{(11)}(\Delta b-\Delta b^{*})(u_{1}-u_{1}^{{\rm st}})
+ \beta^{(02)}(u_{1}-u_{1}^{{\rm st}})^2 \nonumber\\
&+& \beta^{(03)}(u_{1}-u_{1}^{{\rm st}})^3 + ...,
\end{eqnarray}
where $\beta^{(10)}$, $\beta^{(11)}$, $\beta^{(02)}$, $\beta^{(03)}$ are coefficients whose expressions can be derived \footnote{For example, for the polarization transition in Fig. \ref{fig:eigenvalues_pair} we obtain $\beta^{(10)} = 0.226376$, $\beta^{(02)} = 0.672844$, $\beta^{(11)} = 0.789372$, and $\beta^{(03)} = -1.55356$.} from the series expansion of the center manifold \cite{Peralta:2020}.

We consider the most common bifurcation found as a function of the bias asymmetry $\Delta b$ (see Fig. \ref{fig:phases_pair:c} and Fig. \ref{fig:phases_group:c}), i.e. the saddle node bifurcation with $\beta^{(10)} \neq 0$ and $\beta^{(02)} \neq 0$. In its simplified form we have

\begin{equation}
\label{saddle_node_normal}
\frac{d u_{1}}{d t} = \beta^{(10)} (\Delta b-\Delta b^{*}) + \beta^{(02)}(u_{1}-u_{1}^{{\rm st}})^2,
\end{equation}
where the higher order terms can be disregarded. 
Assuming positive coefficients $\beta^{(10)} > 0$, $\beta^{(02)} > 0$, and for $\Delta b < \Delta b^{*}$, the solution of Eq. (\ref{saddle_node_normal}) is

\begin{equation}
\label{sol_c_positive}
u_{1}(t) = u_{1}^{{\rm st}} - \sqrt{\frac{\beta^{(10)} (\Delta b^{*}-\Delta b)}{\beta^{(02)}}} \tanh \left[  \sqrt{\beta^{(10)} \beta^{(02)} (\Delta b^{*}-\Delta b)} t + \mathcal{C} \right],
\end{equation}
while for $\Delta b > \Delta b^{*}$ it is
\begin{equation}
\label{sol_c_negative}
u_{1}(t) = u_{1}^{{\rm st}} + \sqrt{\frac{\beta^{(10)} (\Delta b-\Delta b^{*})}{\beta^{(02)}}} \tan \left[  \sqrt{\beta^{(10)} \beta^{(02)} (\Delta b-\Delta b^{*})} t + \mathcal{C} \right],
\end{equation}
where $\mathcal{C}$ is determined from the initial condition $u_{1}(0)$. Note how, for $\Delta b < \Delta b^{*}$ in Eq. (\ref{sol_c_positive}), the system goes to a stable fixed point in the infinite time limit as $\tanh(\infty)=1$ and $u_{1}(\infty)$ takes a finite value, while for $\Delta b > \Delta b^{*}$ in Eq. (\ref{sol_c_negative}) the system slows down close to $u_{1}(t) \approx u_{1}^{{\rm st}}$ and then diverges (corresponding to a meta-stable state). According to Eq. (\ref{sol_c_negative}), close to the bifurcation point $\Delta b \gtrsim \Delta b^{*}$, the solution scales as $u_{1}(t) - u_{1}^{{\rm st}} \sim (\Delta b-\Delta b^{*})^{1/2}$ and time as $t \sim (\Delta b-\Delta b^{*})^{-1/2}$.

From the previous derivation we infer that, as we increase bias asymmetry $\Delta b$, we find a critical value $\Delta b^{*}$ (see Fig. \ref{fig:phases_pair:c} and Fig. \ref{fig:phases_group:c}) where two fixed points, a saddle and a stable point, merge and disappear at a saddle node bifurcation. At the critical point we observe a meta-stable (slow) region in both variables and time. The size of this region scales with the distance to the critical point as $(\Delta b-\Delta b^{*})^{1/2}$ in the variables and as  $(\Delta b-\Delta b^{*})^{-1/2}$ in time. The zone is centered at the position where the two fixed points merge, i.e. the fixed point at the critical point $\Delta b=\Delta b^{*}$, and it is elongated along the slow eigendirection $\vec{v}_{1}$.

\subsection{Convergence times}

An important quantity to analyze the global dynamical behavior of the model is the convergence time needed to reach the final (stationary) state. Often, when modeling the opinion dynamics of a population, we are not only interested in the possible final states of the system, but also in this convergence time and its dependence on the parameters of the model.

\begin{figure}[ht]
\centering
\subfloat[]{\label{fig:Times_voter:a}\includegraphics[width=0.4\textwidth]{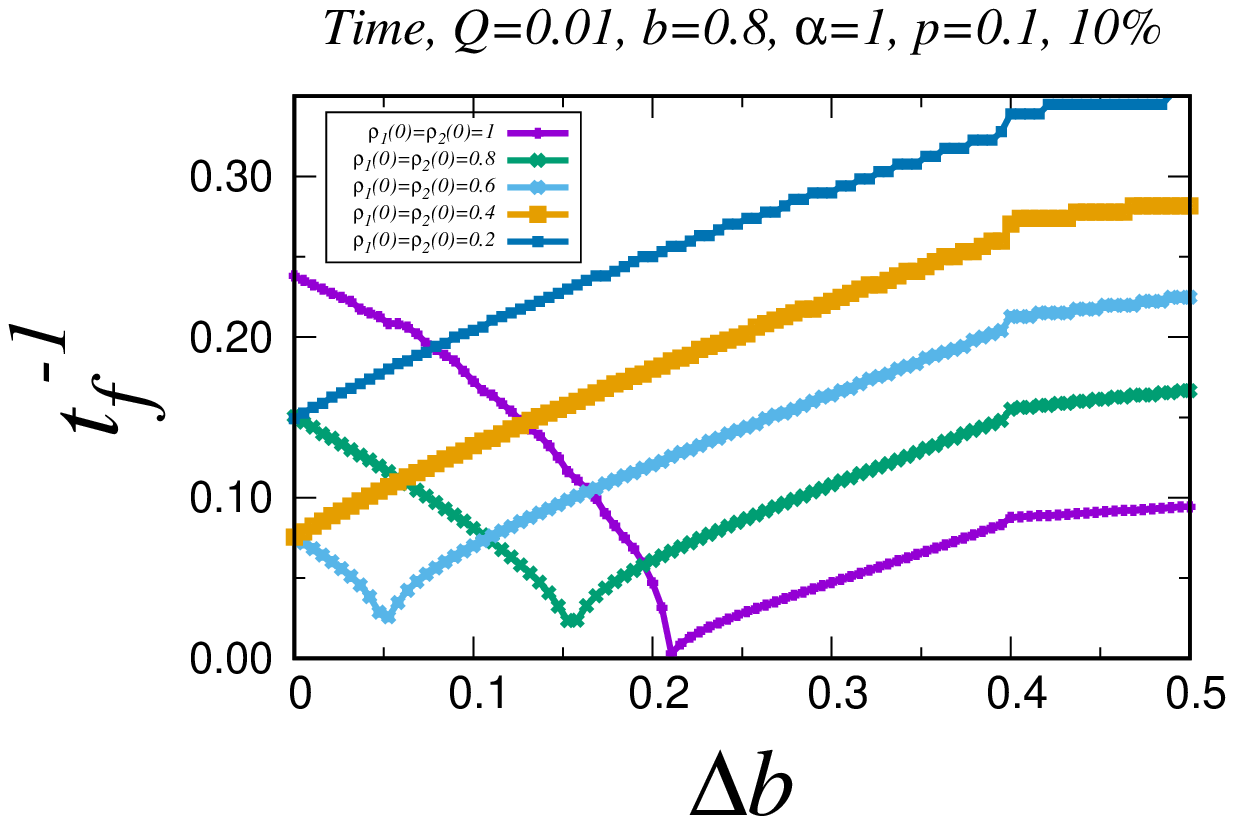}}
\subfloat[]{\label{fig:Times_voter:b}\includegraphics[width=0.4\textwidth]{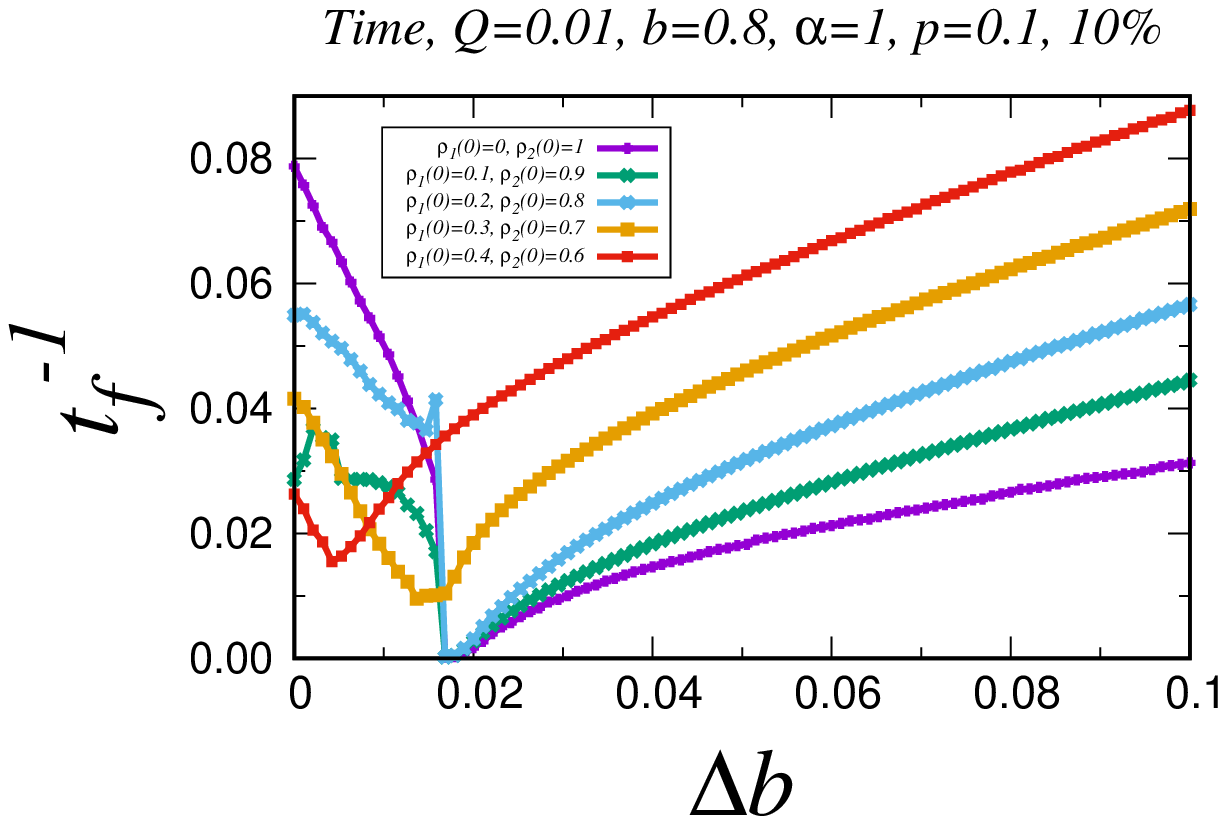}}

\subfloat[]{\label{fig:Times_voter:c}\includegraphics[width=0.4\textwidth]{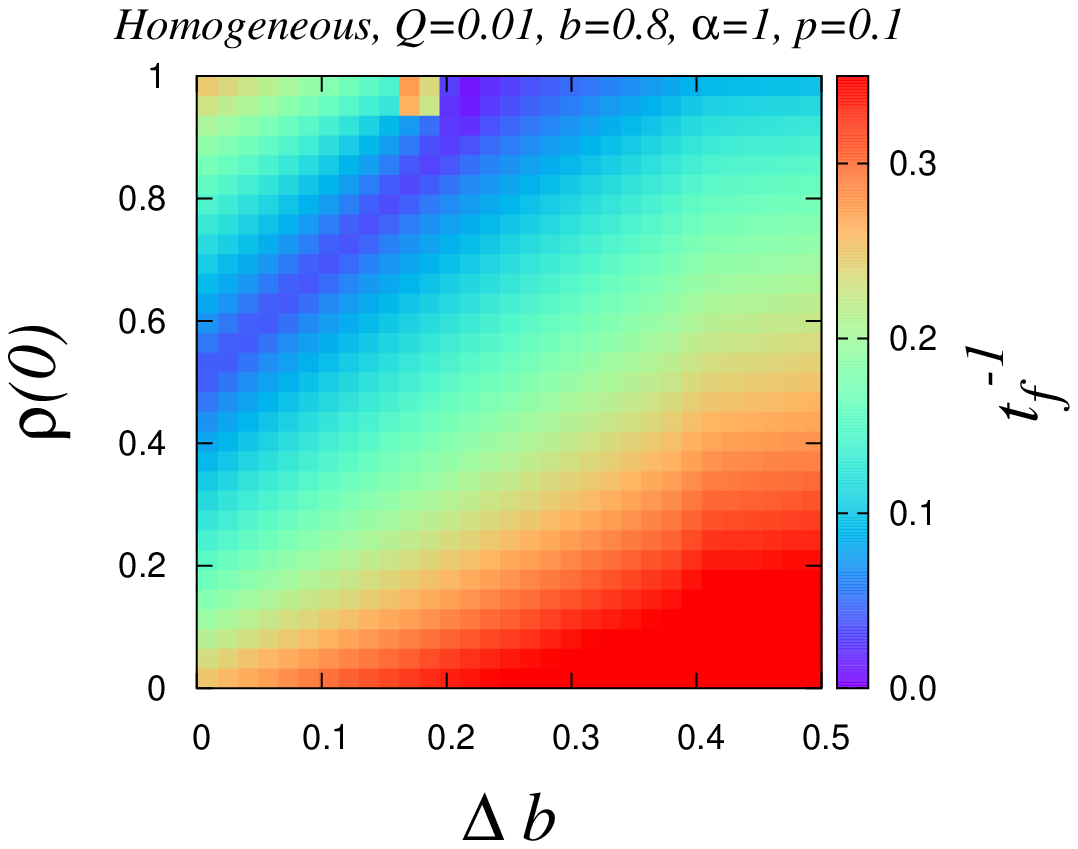}}
\subfloat[]{\label{fig:Times_voter:d}\includegraphics[width=0.4\textwidth]{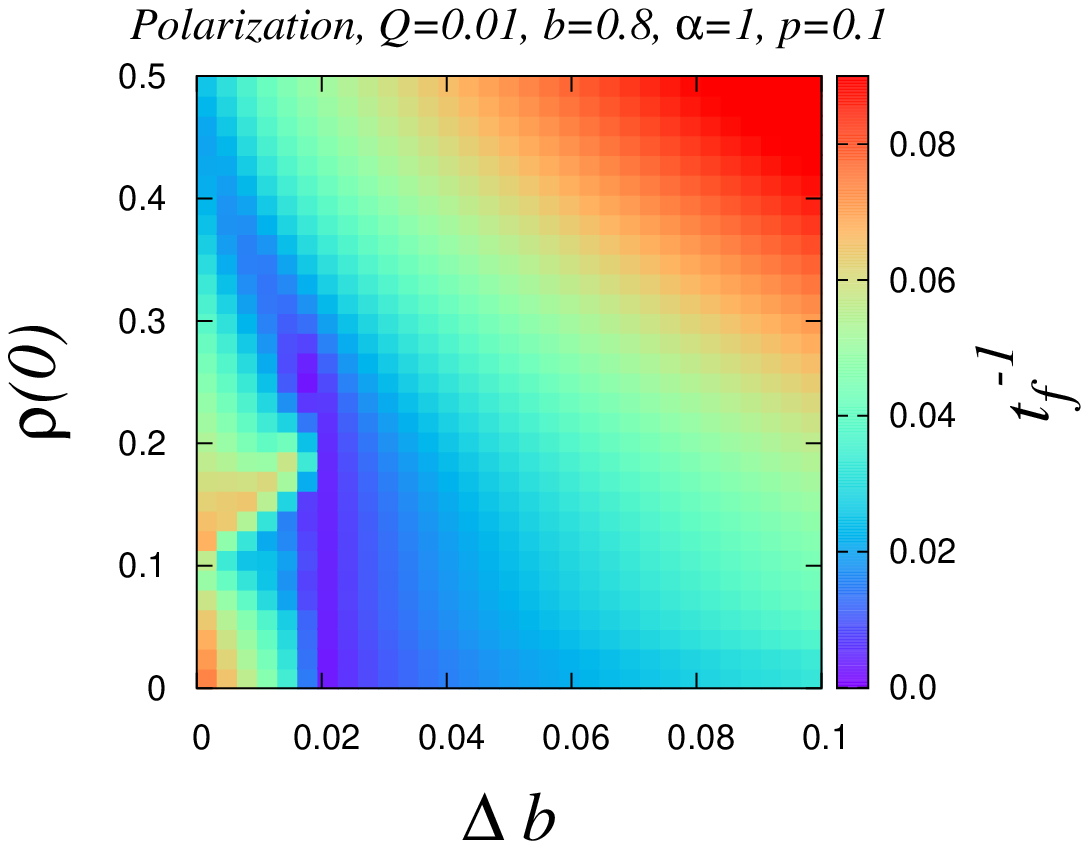}}

\caption{Inverse of convergence time ($t_{f}^{-1}$) as a function of bias asymmetry $\Delta b$. We use the same parameters of the model, networks and simulation details as in Figs. \ref{fig:Trajectories_hom} and \ref{fig:Trajectories_pol}. Top panels show the dependence $t_{f}^{-1}(\Delta b)$ for fixed initial conditions specified in the keys: (a) homogeneous $\rho_{1}(0)=\rho_{2}(0)=\rho(0)$, and (b) polarized $\rho_{2}(0)=1-\rho_{1}(0)=\rho(0)$. Bottom panels show the full dependence $t_{f}^{-1}(\Delta b,\rho(0))$ as a colormap, for (c) homogeneous and (d) polarized initial conditions.}
\label{fig:Times_voter}
\end{figure}

For reason of convenience, we write the time dependence of the global quantities as follows:
\begin{eqnarray}
\label{time_rho1}
\rho_{1}(t) &=& \rho^{{\rm st}}_{1} + (\rho_{1}(0)-\rho^{{\rm st}}_{1}) g_{1}(t),\\
\label{time_rho2}
\rho_{2}(t) &=& \rho^{{\rm st}}_{2} + (\rho_{2}(0)-\rho^{{\rm st}}_{2}) g_{2}(t),
\end{eqnarray}
such that $g_{1,2}(0)=1$ and $g_{1,2}(\infty)=0$. In the linear regime of Eqs. (\ref{rho1_lin}, \ref{rho2_lin}), $g_{1,2}(t)$ are a sum of two exponential functions ($e^{-\lambda_{1} t}$ and $e^{-\lambda_{2} t}$) with amplitudes depending on the eigenvectors and initial conditions. The quantities $\lambda_{1}^{-1}$ and $\lambda_{2}^{-1}$ estimate well the two time scales involved in the time evolution close to the fixed point. For the general global dynamics $g_{1,2}(t)$, the exponential form is not valid and we have a complicated time-dependence. We measure the time scale $t_{f}$ of the global dynamics as the smallest of the solutions $t^{(1)}_{f}$ and $t^{(2)}_{f}$ fulfilling the following conditions:

\begin{eqnarray}
\frac{\rho_{1}(t^{(1)}_{f})-\rho^{{\rm st}}_{1}}{\rho_{1}(0)-\rho^{{\rm st}}_{1}} &=& g_{1}(t^{(1)}_{f}) = \% VAL,\\
\frac{\rho_{2}(t^{(2)}_{f})-\rho^{{\rm st}}_{2}}{\rho_{2}(0)-\rho^{{\rm st}}_{2}} &=& g_{2}(t^{(2)}_{f}) = \% VAL,
\end{eqnarray}
where $\% VAL$ is an arbitrary percentage measuring how close the system is to the final state when the convergence time $t_{f}$ is reached.

In Fig. \ref{fig:Times_voter} we show the dependence of the convergence time $t_{f}$, extracted from numerical simulations, on bias asymmetry and initial conditions in the case of pair interactions, for the same parameters as in Figs. \ref{fig:Trajectories_hom} and \ref{fig:Trajectories_pol}. Note that the system reaches different final states depending on the values of $\Delta b$, $\rho_{1}(0)$ and $\rho_{2}(0)$. This can be clearly seen in Figs. \ref{fig:Times_voter:c} and \ref{fig:Times_voter:d}, where a dark blue line separates two zones where the system reaches different final states, and where the behavior of the convergence times changes (in Figs. \ref{fig:Times_voter:a} and \ref{fig:Times_voter:b} it corresponds to the minimum of the curves). The line that separates the two behaviors is what we call a threshold initial condition $\rho_{0}(\Delta b)$ in Section \ref{sec_num}, i.e. the limit of the basin of attraction of the possible (consensus and polarization) final states. The dependence of the inverse of the convergence time ($t_{f}^{-1}$) with bias asymmetry $\Delta b$, increasing or decreasing, shows a clear correspondence with the eigenvalues of the final (stable) state (polarization or consensus) in Fig. \ref{fig:eigenvalues_pair}. When the final state is polarization ($\Delta b < 0.02$ in Figs. \ref{fig:Times_voter:b} and \ref{fig:Times_voter:d}) it seems that the trend, increasing or decreasing, as a function of $\Delta b$ is not that clear. This can be also understood from Fig. \ref{fig:eigenvalues_pair}, where there is one eigenvalue increasing and another decreasing with $\Delta b$. Generally, the smallest eigenvalue dominates the dynamics, unless the initial condition is aligned with the fast eigendirection. That is the reason why we mainly observe a decreasing behavior in the simulations, while an increasing trend is also possible in some situations. 

In Fig. \ref{fig:Times_voter:a} and \ref{fig:Times_voter:b} we can identify the characteristic scaling behavior $t_{f}^{-1} \sim \vert \Delta b - \Delta b^{*} \vert^{1/2}$, of the saddle-node bifurcation. Before the transition point ($\Delta b \lesssim \Delta b^{*}$), this is directly related to the eigenvalue of the final state $\lambda \sim (\Delta b^{*}-\Delta b)^{1/2}$ and can be considered as critical slowing down, while after the transition ($\Delta b \gtrsim \Delta b^{*}$), the meta-stable state dominates the dynamics (see Fig. \ref{fig:Traj_pol:d}) with an equivalent time scaling as in Eq. (\ref{sol_c_negative}). Note that there are two saddle-node bifurcations, Figs. \ref{fig:Times_voter:b} (consensus) and \ref{fig:Times_voter:a} (polarization), with different transition points $\Delta b^{*}$.

%
%
%

\section{Summary and conclusions}

In this paper we have studied the role of algorithmic bias and community structure in the potential rise of polarization of opinions in online social networks. We have devoted special attention to the temporal behavior of an archetypal two-state opinion-formation model, the language model, as well as to the role of the bias asymmetry $\Delta b$, i.e. the possibility that the online platform favors one opinion over the other. We have derived a pair of mean-field differential equations for the relevant variables of the dynamics, the density $\rho_{1}(t)$ $[\rho_{2}(t)]$ of nodes in group 1 (2) holding opinion $1$. This theoretical description accurately captures the phenomenology of the model and shows a good fit with numerical simulations.

The possible final opinion states reproduced by the model are: {\it consensus} ($\rho_{1} = \rho_{2} \approx 0$ or $\rho_{1} = \rho_{2} \approx 1$), {\it coexistence} ($\rho_{1} = \rho_{2} \approx 1/2$), and {\it polarization} ($\rho_{1} \approx 0$, $\rho_{2} \approx 1$ or $\rho_{1} \approx 1$, $\rho_{2} \approx 0$). All states are found in the whole spectrum between pair and group interactions displayed by the language model as the $\alpha$ parameter is changed. For some parameter values in the group interaction case, we also find additional polarized (polarization-coexistence) states with $\rho_{1} \approx 0$ and $\rho_{2} \approx 1/2$ (and the three equivalent states exchanging the groups $1 \leftrightarrow 2$ and states $0 \leftrightarrow 1$). Using linear stability analysis, we have determined the phase diagrams for pair and group interactions. In general, we find that sufficiently strong asymmetry in the bias is capable to destroy first the stability of the polarized states, and then one of the consensus states via saddle-node bifurcations. The phase diagram of the consensus states corresponds to the well-known cusp catastrophe for pair interactions and butterfly catastrophe for group interactions. Thus, bias asymmetry is a means to `select' final states of the dynamics, controlling the global behavior of the system.

 When the population is divided in two asymmetric groups in terms of size or connectivity (which can be thought of as a majority and minority groups), a somewhat different situation relating to the polarized states is produced by the model. In a range of values of the bias asymmetry $(-\Delta b, \Delta b)$, polarization is not necessarily suppressed but also favored, while above that range it is only destroyed. The two polarized states are not equivalent, depending on which is the opinion of the majority and minority groups. If the social platform benefits the opinion of the minority group then polarization is promoted, while in the opposite case polarization is suppressed in favor of consensus. The values of the bias asymmetry that delimit this behavior can be understood as the case where the structural asymmetry of the network (in group size and connectivity) is compensated by the `favoritism' of algorithmic bias. 
 

Results of numerical simulations confirm the possible final states predicted by the mean-field theory and the role of bias asymmetry. We have found that the convergence time (time to reach a stationary final state) depends on bias asymmetry, initial conditions, and the other parameters of the model in non-trivial ways. By means of the eigenvalues of the linearized dynamics, we were able to characterize this dependence close to the final states, gaining a better understanding of the type of transitions we have, and what happens to the dynamics when one of the final states loses its stability. An unavoidable phenomenon we observe when one of the stable solutions disappears is the presence of meta-stable states. This means that the system becomes trapped for a long period of time in a region of the phase space. Using the normal form of the bifurcations we have derived the scaling relations of the meta-stable zone as a function of the bias asymmetry. This shows a double square root law: $\rho_{1, 2}(t) \sim (t_{f})^{-1}\sim (\Delta b- \Delta b^{*})^{1/2}$ .

In conclusion, we have explored a simplified opinion formation model including some of the arguably most relevant features driving opinion dynamics on online social networks: spreading processes (pair or group copying together with independent behavior of individuals), algorithmic bias, and an underlying networked community structure. The joint effect of these ingredients produces a complex phase space of collective social behavior including coexistence, consensus, and polarization of opinions. We used this formalism as testing ground to study the influence of algorithmic bias on online communication dynamics. We showed that bias imbalance can have a crucial effect on the final opinion state and the dynamics in general. Polarization and consensus states are destroyed for high enough bias asymmetry at different transitions, while just after the transition these destroyed final states become meta-stable. We characterized all possible transitions via phase diagrams. For the local dynamics (close to the final states) we used a linearization of the dynamical equations, while for the global dynamics the normal form of the bifurcation allowed us to detect meta-stable states. Finally, we calculated convergence times both from the theoretical description and by means of numerical simulations.

The aim of such simplified modelling cannot be the quantitative reproduction of some empirical observations, e.g., related to elections. Still, we think that the richness of the behavior of the model, the relatively large number of relevant parameters, and the non-triviality of the temporal evolution of opinions are all features with strong relation to real-world systems. For example, the effect of the parameters on polarization is sometimes counter-intuitive: How algorithmic bias influences the outcome of the dynamics depends strongly on the type of interaction. In our model we can tune interactions from pairwise to group -- in reality both are present and highly heterogeneous. Another relevant observation is the frequent appearance of meta-stable states. As real-world phenomena evolve over finite times (e.g., opinions matter on election day), meta-stable states may be crucial in forecasting efforts. In the future we would like to combine modeling with empirical data analysis, partly from observational data and partly from controlled social experiments, in order to further understand the interplay between human collective action and online algorithms.


\section*{Acknowledgements}

We thank Matteo Neri for his help in the initial stages of this project. We acknowledge support from AFOSR (Grant \#FA8655-20-1-7020). J.K. is grateful for support by project ERC DYNASNET Synergy (Grant \#810115). J.K. and G.I. acknowledge support from projects EU H2020 Humane AI-net (Grant \#952026) and SAI enabled by FWF (I 5205-N) within the EU CHIST-ERA program (call 2019).

\section*{References}

\bibliographystyle{unsrt}

\end{document}